\documentclass[trackchanges,twocolumn,twocolappendix]{aastex701}
\usepackage{CJK}
\usepackage{amsmath}

\usepackage{color}

\begin{document}
\begin{CJK*}{UTF8}{gbsn}
\title{A Self-Consistent Jeans Analysis of the Milky Way Rotation Curve}

\author[orcid=0000-0003-2637-277X,sname='Jiao']{Yongjun Jiao (焦永俊)}
\affiliation{Department of Astronomy, Tsinghua University, Beijing 100084, People’s Republic of China}
\email[show]{jiaoyongjun@tsinghua.edu.cn}  

\author[orcid=0000-0002-2165-5044,sname='Hammer']{Francois Hammer} 
\affiliation{LIRA, Observatoire de Paris, Universit\'e PSL, CNRS, Meudon 92195, France}
\email[show]{francois.hammer@obspm.fr}

\author[orcid=0000-0002-1778-7888,sname=Akib]{Istiak Akib}
\affiliation{LIRA, Observatoire de Paris, Universit\'e PSL, CNRS, Meudon 92195, France}
\affiliation{Center for Astronomy, Space Science and Astrophysics, Independent University, Bangladesh, Dhaka 1229, Bangladesh}
\email{Istiak-Hossain.AKIB@obspm.fr}

\author[orcid=0000-0001-7949-3407]{Yanbin Yang (杨雁宾)}
\affiliation{LIRA, Observatoire de Paris, Universit\'e PSL, CNRS, Meudon 92195, France}
\email{yanbin.yang@obspm.fr}

\author[orcid=0000-0001-8459-1036]{Haifeng Wang (王海峰)}
\affiliation{Local Universe and Time-Domain Astronomy Laboratory, LuoxiaHong Institute of Astronomy, China West Normal University, Nanchong 637002, People's Republic of China}
\email{haifeng.wang.astro@gmail.com}





\begin{abstract}

Jeans-based measurements of the Milky Way (MW) rotation curve (RC) have revealed a declining outer RC. The interpretation, however, depends sensitively on the tracer selection, the tracer density profile, gradient terms in the Jeans equation, and non-axisymmetric perturbations in the outer disk. We present a more self-consistent Jeans analysis of the MW RC and reassess whether the outer decline remains after a revised treatment of the main Jeans terms and their associated systematic uncertainties, including those related to non-axisymmetric perturbations. We select a kinematically homogeneous disk tracer sample designed to retain stars on nearly circular disk orbits, removing only 12\% of the original sample. We estimate the tracer density profile from the same stars used for the kinematic moments, rather than adopting an external scale length. We find that the tracer density profile is better described by a double exponential over the validated range $12.5<R<21\,\mathrm{kpc}$. We also assess the main systematic uncertainties associated with the fiducial Jeans equation terms, including the often neglected cross term, and account for different rotational velocities in the outer disk substructures. The derived RC remains nearly flat in the inner disk and declines at large radii. Over the adopted outer interval, $R\gtrsim16\,\mathrm{kpc}$, the RC slope is consistent with a Keplerian decline even after accounting for the systematic uncertainties. The corresponding mass modeling gives an extrapolated dynamical mass of $2.01^{+0.10}_{-0.08}\times10^{11}\,M_\odot$.

\end{abstract}



\section{Introduction}
\label{sec:introduction}

The existence of dark matter (DM) was first proposed by \citet{Zwicky1933}, who found an unexpectedly high velocity dispersion of galaxies within the Coma Cluster. In the 1970s, the discovery of flat rotation curves (RCs) in spiral galaxies further supported the idea that DM constitutes the dominant mass component in galaxies \citep{Rubin1978,Bosma1978}. RCs remain one of the most robust methods for probing the DM distribution in disk galaxies. However, our location within the disk of the Milky Way (MW) has historically limited our ability to construct a precise RC and thus to constrain its mass distribution.

A significant breakthrough came with \textit{Gaia}, whose astrometry has enabled three-dimensional velocity measurements of individual stars. Based on its second data release (DR2), \citet{Eilers2019} provided the first evidence for a declining RC in the MW by applying spectrophotometric parallax distances from \citet{Hogg2019} to a sample of $\sim$23,000 red giant branch (RGB) stars. Similar declining behavior has been confirmed by several independent studies based on \textit{Gaia} Data Release 3 \citep[DR3,][]{Wang2023,Ou2024,Jiao2023}.

In particular, \citet{Jiao2023} reported an outer MW RC slope consistent with a Keplerian decline beyond $R \gtrsim 19\,\mathrm{kpc}$. This observed Keplerian decline, combined with the relatively low MW dynamical mass, poses a challenge to the standard Lambda cold dark matter ($\Lambda$CDM) cosmological framework.
First, the decline differs from the nearly flat RCs often observed in external disk galaxies, although most such RCs are measured
using gaseous tracers, principally H\,I and H$\alpha$. It begins near the optical disk radius ($\sim$17 kpc), suggesting that the MW halo is more compact than previously estimated by \citet{Bland-Hawthorn2016} and than expected from the CDM paradigm \citep{Li2025}. Second, the baryonic component, including stars, dust, and cold gas, is estimated to be $0.6 \times 10^{11}\,\text{M}_\odot$, or nearly 30\% of the total dynamical mass \citep[$2.06 \times 10^{11}\,\text{M}_\odot$ within $\sim120$ kpc,][]{Jiao2023}. This baryon fraction exceeds both typical values in other spiral galaxies ($\sim$10\%) and the cosmic average \citep[$\sim$17\%,][]{Planck2020}.

Despite these findings,  the maximum MW mass consistent with \textit{Gaia} DR3 rotation curves has not yet been evaluated with a careful treatment of the main systematic uncertainties. \citet{Jiao2023} included most of the systematic uncertainties identified by \citet{Eilers2019}, but noted that this procedure can overestimate them. \citet{Ou2025} performed a comprehensive evaluation of potential systematic uncertainties from a systematic comparison with FIRE simulations, but they likely overestimate them because the comparison models have significantly higher rotation velocities at all radii than those observed \citep[see models m2f, m2i and m2m in their Figure B3]{Hammer2024Simu}. This leads to a rather large MW mass range of $2$--$8 \times 10^{11}\,\mathrm{M}_\odot$, comparable to earlier \textit{Gaia} DR2 estimates \citep{Jiao2021}, which casts some doubt about the impact of the improved precision of DR3.
 
Beyond 22 kpc, the systematic uncertainties are dominated by the non-axisymmetric terms in the Jeans equation, such as those caused by warps and flares, which are supposed to be accounted for using the data splitting methodology \citep{Eilers2019,Jiao2023,Ou2024}. A careful evaluation of systematic uncertainties is therefore essential to understand their origin and limit.

Recent studies have also questioned the reliability of MW RCs based on \textit{Gaia}, especially in the outer disk. For example, \citet{Ibata2024} analyzed stellar streams and found a rotation curve that significantly deviates from that of \citet{Eilers2019} at $R > 15\,\mathrm{kpc}$. \citet{Koop2024} argued that the apparent Keplerian decline may arise from adopting an incorrect stellar density profile in the Jeans analysis. They showed that, if the MW disk is truncated, adopting a pure exponential profile can underestimate the circular velocity by $\Delta V_\mathrm{c} \sim 10$--$45\,\mathrm{km\,s^{-1}}$ at $R = 30\,\mathrm{kpc}$, depending on the truncation steepness ($n = 2$--$4$ in their parameterisation). Such biases could transform the apparent Keplerian decline reported by \citet{Jiao2023} into a more conventional slightly decreasing rotation curve. 

These results show that Jeans-based RC measurements are sensitive to the assumed radial density profile, in particular to the disk scale length $h_R$ and any truncation radius. Rather than adopting a fixed exponential profile or assuming no truncation, the tracer density profile should be estimated from the same data used for the RC. We therefore focus on this source of systematics by determining the radial density distribution of the RGB sample of \citet{Ou2024}. This reduces the bias and uncertainty associated with the assumed disk profile.

This paper reanalyzes the MW RC using the stellar sample of \citet{Ou2024}. We focus on measuring the tracer density profile from the same stars used for the kinematic analysis and on estimating systematic uncertainties without redundancy. The paper is organized as follows. Section~\ref{sec:Methodology} outlines the Jeans framework. Section~\ref{sec:data} describes our data sample. Section~\ref{sec:Measurements} presents the radial binning, tracer density profile, kinematic gradient terms, cross term, and systematic uncertainties. Section~\ref{sec:results} presents the RC and mass modeling. Section~\ref{sec:discussion} compares our findings with previous estimates and discusses the implications, and Section~\ref{sec:conclusions} summarizes our conclusions.

\section{Methodology}
\label{sec:Methodology}
\subsection{Radial Jeans equation}
\label{sec:jeans_framework}

Following recent MW RC studies based on \textit{Gaia} data \citep[e.g.][]{Eilers2019,Wang2023,Zhou2023,Ou2024}, we derive the MW rotation curve from the radial Jeans equation in axisymmetric cylindrical coordinates $(R,\phi,z)$ \citep{Binney2008}. The selected tracers are rotationally supported disk stars (see Section~\ref{sec:data}), for which cylindrical coordinates provide the most natural representation of their density and velocity moments. For an axisymmetric stellar system, the radial Jeans equation is
\begin{equation}
\frac{\partial \nu \langle v_R^2\rangle}{\partial R}
+\frac{\partial \nu \langle v_R v_z\rangle}{\partial z}
+\nu \left(
\frac{\langle v_R^2\rangle-\langle v_\phi^2\rangle}{R}
+\frac{\partial \Phi}{\partial R}
\right)=0,
\label{eq:jeans_radial}
\end{equation}
where $\nu(R,z)$ is the tracer density and $\Phi$ is the Galactic potential. Assuming a steady state and reflection symmetry about the Galactic mid-plane, such that $\langle v_Rv_z\rangle=0$ at $z=0$, and evaluating the equation at the disk mid-plane, the circular velocity can be written as
\begin{equation}
V_{\rm c}^2(R)=
\langle v_\phi^2\rangle
-\langle v_R^2\rangle
\left[
1+\frac{\partial \ln \nu}{\partial \ln R}
+\frac{\partial \ln \langle v_R^2\rangle}{\partial \ln R}
\right]
- R\,\frac{\partial \langle v_R v_z\rangle}{\partial z}.
\label{eq:jeans_vc_general}
\end{equation}

The quantities entering Eq.~(\ref{eq:jeans_vc_general}) are directly measured kinematic moments and gradients of the tracer and kinematic profiles. Following \citet{Eilers2019}, we estimate the kinematic second moments from the velocity moment tensor,
\begin{equation}
\mathbf{V}\leftarrow \langle \mathbf{v}\mathbf{v}^{\rm T}\rangle - \mathbf{C}_{\mathbf v},
\label{eq:velocity_tensor}
\end{equation}
where $\mathbf{C}_{\mathbf v}$ denotes the covariance matrix of the velocity measurement uncertainties. The diagonal components $V_{\phi\phi}$ and $V_{RR}$ estimate $\langle v_\phi^2\rangle$ and $\langle v_R^2\rangle$, respectively, while the off-diagonal component $V_{Rz}$ estimates $\langle v_R v_z\rangle$. We therefore evaluate the cross term in Eq.~(\ref{eq:jeans_vc_general}) from the vertical gradient of $V_{Rz}$. In practice, this gradient is estimated from the variation of $V_{Rz}$ across subsamples at different heights above and below the Galactic plane. The radial gradient $\partial \ln \langle v_R^2\rangle/\partial \ln R$ is obtained from the radial variation of $V_{RR}$, whereas the density gradient $\partial \ln \nu/\partial \ln R$ must be determined from the radial density profile of the same tracer population used in the Jeans analysis.

The derivation and use of Eq.~(\ref{eq:jeans_vc_general}) rely on approximate axisymmetry and dynamical equilibrium. These assumptions are common to recent MW RC studies \citep{Eilers2019,Wang2023,Jiao2023,Ou2024}, and their limitations have been examined explicitly in recent assessments of the robustness of Jeans-based estimates \citep{Koop2024,Ou2025}. In particular, \citet{Ou2025} identified four major sources of bias and uncertainty in Milky Way mass estimates: selection biased tracer populations, inaccurate treatment of the asymmetric drift correction, departures from axisymmetry, and dynamical disequilibrium. We therefore evaluate all terms entering Eq.~(\ref{eq:jeans_vc_general}) as self consistently as possible by optimizing the tracer selection, deriving the density term from the same stars used for the kinematic analysis, explicitly including the cross term, and propagating the residual effects of asymmetry and non-equilibrium into the systematic uncertainty estimation. The detailed implementation is presented in Section~\ref{sec:Measurements}.

\subsection{Radial binning}
\label{sec:binning_requirement}

The choice of radial bins is important for RC measurements, because neighbouring points are not necessarily statistically independent. \citet{Posti2022} showed that neglecting covariance between adjacent radii can lead to overly optimistic constraints in mass modeling. For the MW, \citet{Oman2024} found that this effect is non negligible for recent RC measurements based on \textit{Gaia}. This is part of the broader need to reassess methodological assumptions in Jeans-based MW RC estimates \citep{Koop2024,Ou2025}.

\begin{figure}[htb!]
\centering
\includegraphics[width=\columnwidth]{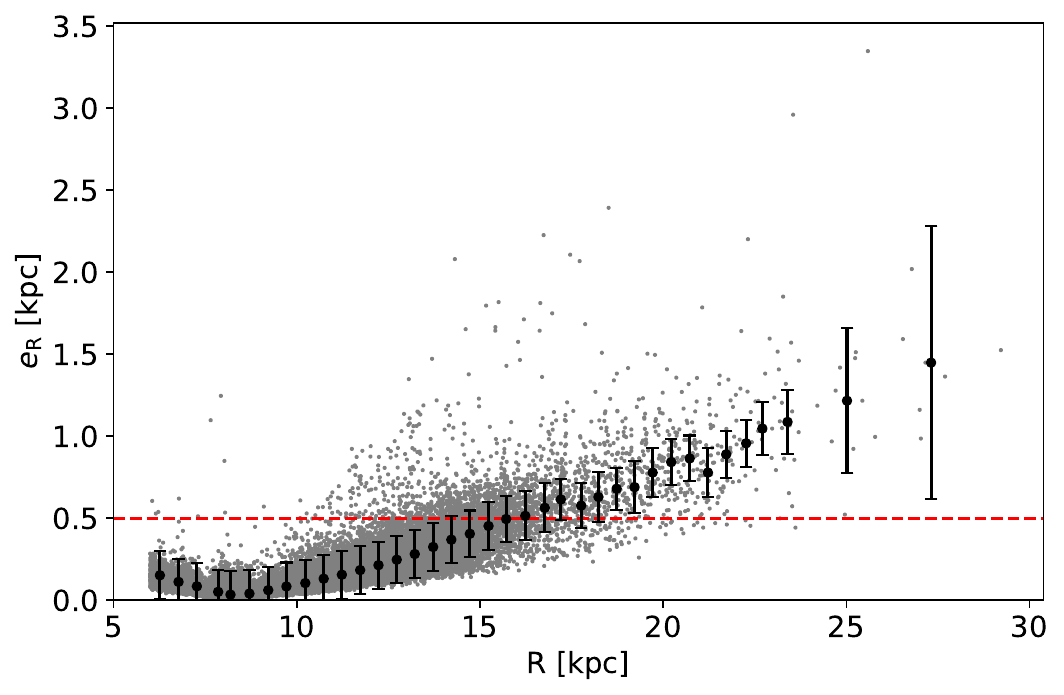}
\caption{Median uncertainty in Galactocentric radius as a function of Galactocentric radius for the sample of \citet{Ou2024}. Black points show the median values in radial bins, and the vertical bars indicate the standard error. The dashed red line marks 0.5 kpc. Beyond $R\sim15\,\mathrm{kpc}$, the median uncertainty exceeds the $0.5\,\mathrm{kpc}$ bin width adopted by \citet{Ou2024}.}
\label{fig:eRGC}
\end{figure}

This point is relevant for the RC of \citet{Ou2024}, which was sampled with radial bins of width $\Delta R=0.5\,\mathrm{kpc}$, except for the outermost three bins. Such fine sampling is appropriate when the radial precision is sufficiently high. However, Fig.~\ref{fig:eRGC} shows that, for their sample, the median uncertainty in Galactocentric radius exceeds $0.5\,\mathrm{kpc}$ beyond $R\sim15\,\mathrm{kpc}$. In the outer disk, the nominal bin width is therefore smaller than the typical radial precision of the data.

We therefore adopt a wider characteristic radial scale of $1\,\mathrm{kpc}$ over the intermediate and outer disk. This scale better matches the effective radial precision of the data, while remaining sufficient to trace the global variation of the RC.

\subsection{Tracer density gradient}
\label{sec:density_gradient}

The tracer density gradient, $\partial \ln \nu / \partial \ln R$, is an important term of Eq.~(\ref{eq:jeans_vc_general}). In many recent Jeans-based measurements of the MW RC, this term is approximated by assuming a single exponential radial profile for the tracer population \citep[e.g.,][]{Eilers2019,Wang2023,Jiao2023,Ou2024},
\begin{equation}
\nu(R)=\nu_0 \exp\left(-\frac{R}{h_R}\right),
\label{eq:single_exp_density}
\end{equation}
where $h_R$ is the radial scale length. Under this assumption,
\begin{equation}
\frac{\partial \ln \nu}{\partial \ln R}=-\frac{R}{h_R},
\end{equation}
and Eq.~(\ref{eq:jeans_vc_general}) becomes
\begin{equation}
V_{\rm c}^2(R)=
\langle v_\phi^2\rangle
-\langle v_R^2\rangle
\left[
1-\frac{R}{h_R}
+\frac{\partial \ln \langle v_R^2\rangle}{\partial \ln R}
\right]
- R\,\frac{\partial \langle v_R v_z\rangle}{\partial z}.
\label{eq:jeans_single_exp}
\end{equation}
This parametrization is convenient, but it shows that the inferred rotation curve depends directly on the adopted tracer scale length.

The difficulty is that the density term entering the Jeans equation should describe the intrinsic three dimensional number density of the same tracer population used to measure the kinematic moments, up to an arbitrary normalization \citep[e.g.,][]{Zhang2013,Bovy2013}. For a spectroscopic survey such as APOGEE, the observed counts are affected by target selection and magnitude limits. A full recovery of the intrinsic tracer density would therefore require a detailed survey selection function. We do not attempt a full reconstruction here. Instead, since the Jeans equation depends on the logarithmic radial slope rather than the absolute normalization, we aim to estimate this slope over a radial range where residual selection effects vary slowly with radius.

We therefore do not adopt a fixed literature value for the tracer scale length. Instead, we infer the radial shape from the same optimized RGB sample used for the RC measurement. We use a $V_{\max}$ corrected number density profile to mitigate the dominant magnitude volume bias, and we restrict the fit to the radial interval where the extinction corrected RGB luminosity functions are most internally consistent. This procedure does not recover the absolute intrinsic RGB density, nor does it replace a full APOGEE selection function. It provides a practical approximation to the tracer density gradient over the validated radial range. This approximation is supported by the good agreement with the independently derived outer disk RGB density profile of \citet{Wang2018}, and by recent work identifying selection biased tracer populations as a major source of bias in Jeans-based MW RC measurements \citep{Ou2025}. The corresponding validation tests, density measurements, and calculation of the density gradient are presented in Section~\ref{sec:density_profile}.

\section{Data sample}
\label{sec:data}

We adopt the stellar sample construction of \citet{Ou2024}, which is based on RGB stars selected from \textit{Gaia} DR3 following the method of \citet{Eilers2019}. Reproducing their selection yields 33,352 RGB stars in our catalogue, 17 more than the 33,335 stars quoted by \citet{Ou2024}. We verified that these 17 additional stars are all located within $R<15\,\mathrm{kpc}$. Because our subsequent analysis further refines the sample, this small difference has negligible impact on the derived MW rotation curve.

To obtain a kinematically homogeneous tracer population suitable for axisymmetric Jeans equation analysis, we apply three additional selection criteria.

\begin{figure}[htb!]
  \centering
  \includegraphics[width=\columnwidth]{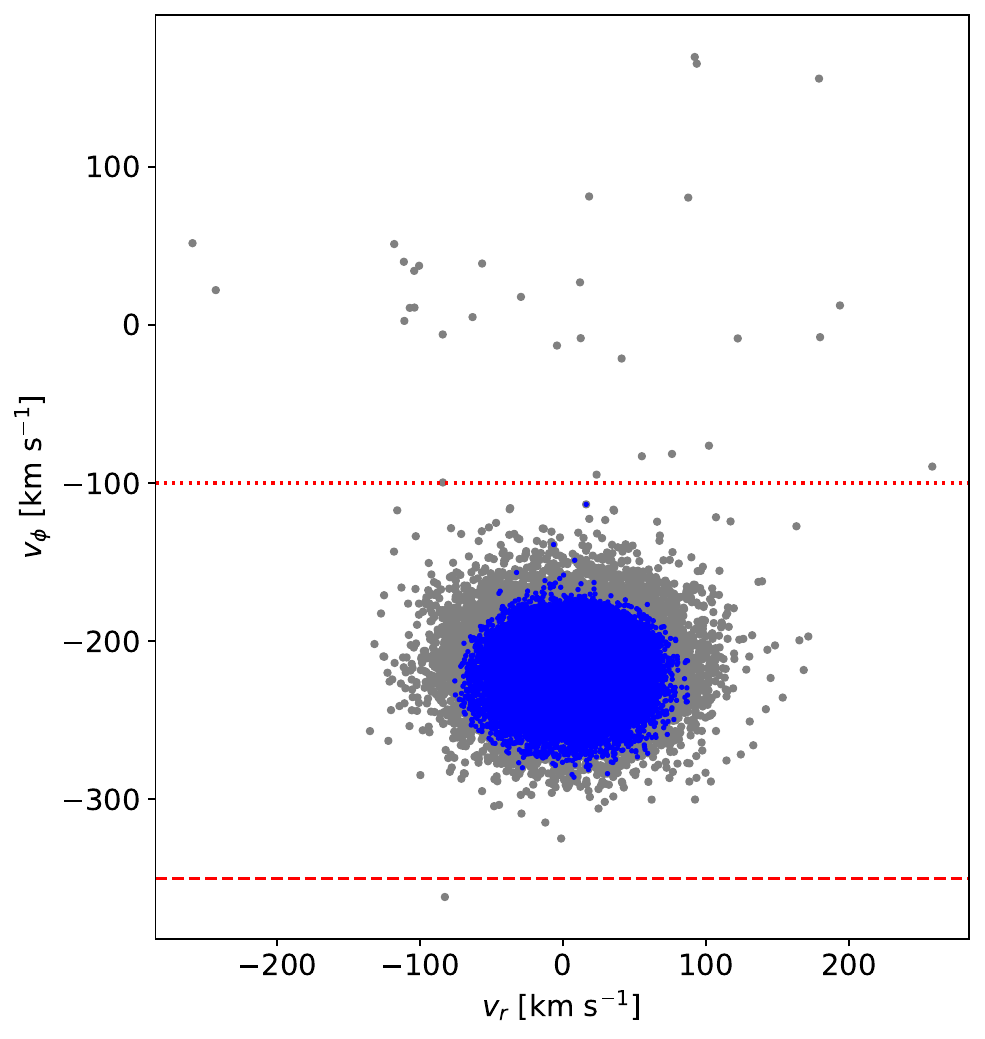}
  \caption{Tangential velocity $v_{\phi}$ versus radial velocity $v_{\rm r}$ for the stellar sample. The original sample from \citet{Ou2024} is shown in gray, while the stars selected in this study are shown in blue. Stars above the horizontal dotted line at $v_\phi = -100~\text{km}\,\text{s}^{-1}$ have weak or inverted rotation, while stars below the dashed line at $v_\phi = -350~\text{km}\,\text{s}^{-1}$ have excessively large prograde tangential velocities. These stars are excluded from our final sample.}
  \label{fig:Ch2_vphi_excluded}
\end{figure}

First, we restrict the azimuthal velocity to $-350 < v_\phi < -100\,\mathrm{km\,s^{-1}}$, where $v_\phi$ is defined in Galactocentric coordinates. This cut excludes counter rotating or weakly rotating stars ($v_\phi > -100\,\mathrm{km\,s^{-1}}$), as well as stars with very large prograde azimuthal velocities ($v_\phi < -350\,\mathrm{km\,s^{-1}}$). Such stars are more likely associated with the thick disk, stellar halo, or other kinematically distinct structures related to disequilibrium processes, such as tidal streams \citep{Ibata1994} or the Gaia-Sausage-Enceladus (GSE) merger \citep{Helmi2018,Haywood2018}. Applied to the reproduced \citet{Ou2024} sample, this cut removes 32 stars, including 3 stars at $R>20\,\mathrm{kpc}$. We verified that the three stars excluded by this kinematic cut correspond to the highly eccentric outer disk stars identified by \citet[see their Fig.~A3]{Ou2024}.

Second, we adopt a vertical selection of $|z| < 2\,\mathrm{kpc}$, where $z$ is the height above the Galactic midplane. At large Galactocentric radii, this selection does not significantly reduce the number of available tracers, because the \citet{Eilers2019} and \citet{Ou2024} samples already follow a wedge geometry in the outer disk (see Fig.~A1 of \citealt{Ou2024}). We use the same geometry here for consistency. A much thinner selection close to the midplane, for example $|z| < 1\,\mathrm{kpc}$, would be more sensitive to the detailed shape of the warped and flaring disk, and could therefore introduce larger systematic shifts in the inferred kinematics. The choice of $|z| < 2\,\mathrm{kpc}$ is a compromise: it preserves the outer disk coverage of the original wedge sample while reducing sensitivity to the assumed midplane of the warped and flaring disk. This relatively broad vertical selection preserves the outer disk coverage and increases the candidate sample to 36,543 stars before the final outlier clipping. In Appendix~\ref{app:systematic_z} we compare our results with those obtained from a subsample restricted to $|z| < 1\,\mathrm{kpc}$ and show that the main conclusions are unchanged.

Third, we remove strong kinematic outliers relative to the bulk of the disk population. We assume that, within a given Galactocentric radial bin, the disk population is approximately described by a bivariate normal velocity distribution. In each bin, we construct the two dimensional velocity distributions in $(v_r, v_\phi)$ and $(v_r, v_z)$, and determine their median values and corresponding $1\sigma$ dispersions. Stars far from the main distribution in either projection are more likely to have been strongly perturbed or to belong to non disk components, rather than tracing the smooth disk potential. After this clipping procedure, the final sample used to derive the RC contains 29,285 stars and is dominated by rotationally supported disk stars.

All other selection criteria follow those of \citet{Eilers2019} and \citet{Ou2024}. Figure~\ref{fig:Ch2_vphi_excluded} compares the $v_\phi$--$v_r$ distribution of the final sample used in this study with that of \citet{Ou2024}.

To transform the observed heliocentric positions and velocities to the Galactocentric frame, we adopt the solar parameters from \citet{Ou2024} for consistency with their analysis. Specifically, we use a solar Galactocentric distance of $R_\odot = 8.178\,\mathrm{kpc}$ \citep{Gravity2019Dis} and a vertical offset of $z_\odot = 20.8\,\mathrm{pc}$ \citep{Bennett2019}. The solar motion with respect to the Galactic centre is taken as $(v_{x,\odot}, v_{y,\odot}, v_{z,\odot}) = (5.1, 247.3, 7.8)\,\mathrm{km\,s^{-1}}$ \citep{Reid2004,Schonrich2010}.

\section{Measurements and systematic uncertainties}
\label{sec:Measurements}

\subsection{Adopted radial bins}
\label{sec:binning}

As discussed in Section~\ref{sec:binning_requirement}, the radial sampling of the RC should follow the effective radial precision of the data, rather than the nominal bin width alone. We therefore revise the binning used in previous work and adopt a characteristic radial scale of $1\,\mathrm{kpc}$ over most of the RC analysis.

\begin{figure}[htb!]
\centering
\includegraphics[width=\columnwidth]{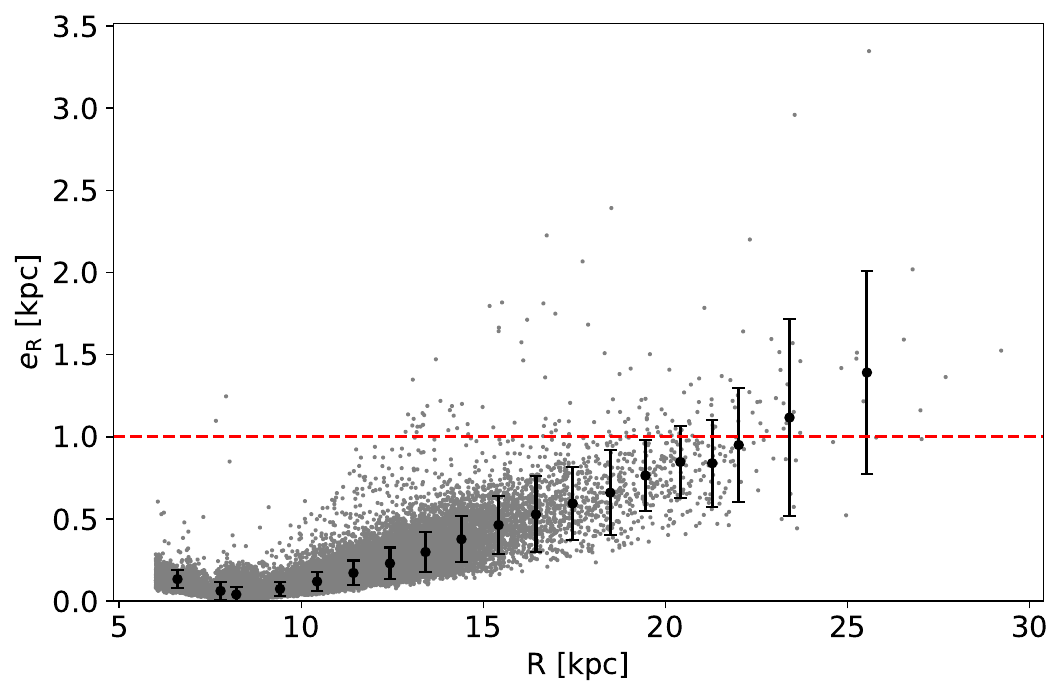}
\caption{Median uncertainty in Galactocentric radius as a function of Galactocentric radius for the sample used in this study. Gray points show individual stars, while black points indicate the median uncertainty in each radial bin. The dashed red line marks $1\,\mathrm{kpc}$. Over most of the radial range used in the RC analysis, the median uncertainty is comparable to or below this scale. The outermost bins are set by the requirement of a minimum number of stars.}
\label{fig:eRGC_new}
\end{figure}

Figure~\ref{fig:eRGC_new} shows the uncertainty in Galactocentric radius for the present sample. Over most of the radial range used for the RC measurement, the median uncertainty is comparable to or below $1\,\mathrm{kpc}$. This scale therefore better matches the data than the finer binning used previously. In the outer disk, however, the number of tracers decreases rapidly. We therefore do not define the outermost bins by radial precision alone. Instead, we require at least 16 stars per bin to obtain statistically meaningful estimates of the kinematic moments. This binning strategy retains the global radial behavior of the RC while reducing oversampling at intermediate and large radii.

\subsection{Tracer density profile}
\label{sec:density_profile}

\begin{figure*}[htb!]
  \centering
  \includegraphics[width=0.9\textwidth]{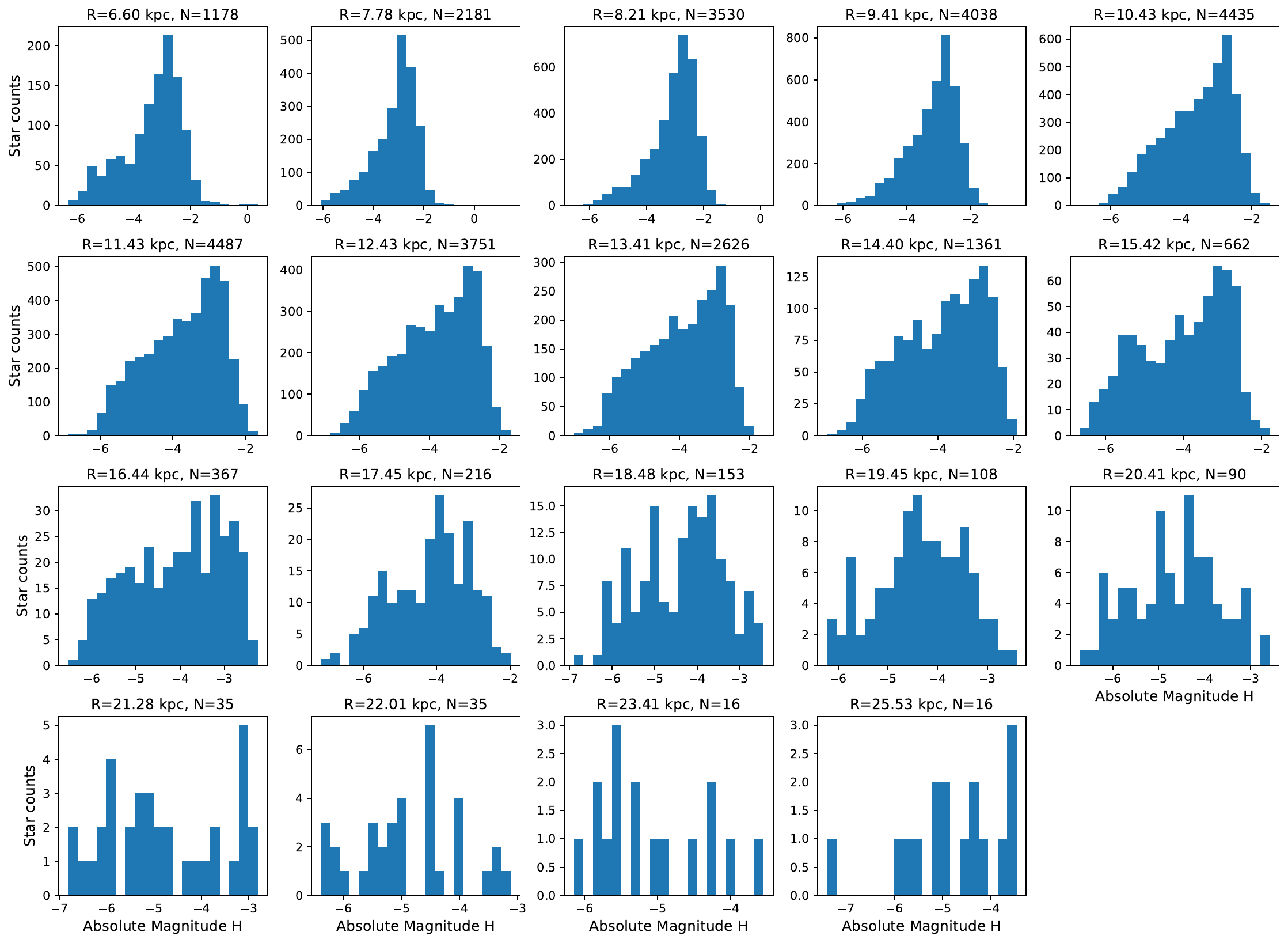}
  \caption{Distribution of extinction corrected absolute $H$ band magnitude in each radial bin. The extinction correction uses the three dimensional dust map of \citet{Green2019}. The title of each panel gives the bin centre and the number of stars. The intermediate bins show broadly similar luminosity function shapes, while the inner bins are more affected by the bright magnitude limit and the outer bins by small number statistics.}
  \label{fig:h_abs_bins}
\end{figure*}

Following \citet{Eilers2019} and \citet{Ou2024}, our RC measurement is based on low-$\alpha$ upper RGB stars, which mainly trace a young or intermediate age disk population (smaller than 7 Gyr old, \citealt{Haywood2019}). The tracer density term in Eq.~(\ref{eq:jeans_vc_general}) should refer to the same population used to measure the kinematic moments. We therefore estimate the radial density profile directly from the RC tracer sample.

For each star, we compute a $V_{\max}$ correction from the APOGEE cohort dependent $H_{\min}$ and $H_{\max}$ limits \citep{Zasowski2017}. We then use these weights to derive a number density estimate in each radial bin. Since the Jeans equation depends only on the logarithmic density gradient, the absolute normalization is irrelevant. The associated uncertainties are estimated assuming Poisson statistics for the weighted counts. This correction mitigates the dominant magnitude volume bias, but it is not a full reconstruction of the APOGEE selection function \citep{Majewski2017,Zasowski2017}. We therefore fit the density profile only over the radial interval where the observed RGB luminosity function is most stable with radius.

\begin{figure}[htb!]
  \centering
  \includegraphics[width=\columnwidth]{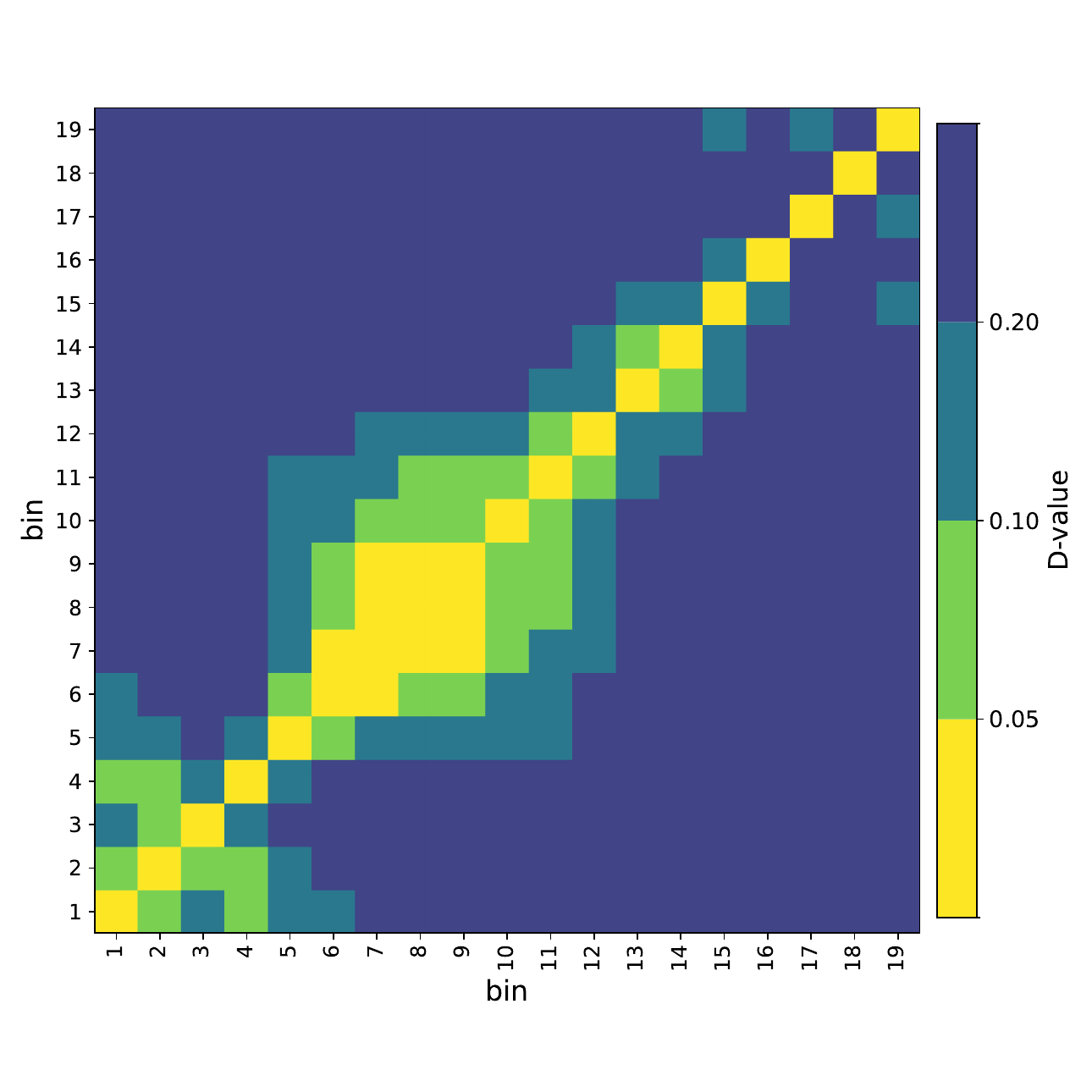}
  \caption{Pairwise KS $D$ statistic for the $H_{\rm abs}$ distributions in different radial bins (see Fig.~\ref{fig:h_abs_bins}). Lower $D$ values indicate more similar distribution shapes. The bins adopted for the density fit form the broadest contiguous region with low $D$ values.}
  \label{fig:ks_pair}
\end{figure}

\subsubsection{Validation of the fitted radial range}
\label{sec:density_range}

As discussed in Section~\ref{sec:density_gradient}, the density gradient entering the Jeans equation should be measured over a range where residual selection effects vary as little as possible with radius. We test this using the extinction corrected absolute $H$ band magnitude, $H_{\rm abs}$, of the selected stars after correcting for extinction with the three dimensional dust map of \citet{Green2019}. If different radial bins sample different parts of the RGB luminosity function, the inferred density slope can be biased even after the $V_{\max}$ correction.

Figure~\ref{fig:h_abs_bins} shows the $H_{\rm abs}$ distribution in each radial bin. The inner bins change systematically with radius, as expected when the APOGEE bright limit removes nearby luminous RGB stars. The outermost bins contain too few stars to define a stable luminosity function and are more susceptible to the loss of intrinsically fainter tracers. By contrast, the intermediate bins from roughly $R=12.5$ to $21\,\mathrm{kpc}$ show more similar $H_{\rm abs}$ distributions. They therefore provide the most suitable range for estimating the radial number density profile.

We quantify this comparison with pairwise two sample Kolmogorov--Smirnov (KS) tests between the $H_{\rm abs}$ distributions of all radial bins. We use the KS $D$ statistic as the main diagnostic, because it directly measures the difference in distribution shape. As shown in Fig.~\ref{fig:ks_pair}, the bins centred between $R=12.43$ and $20.41\,\mathrm{kpc}$ form the broadest contiguous region of mutually low $D$ values. The corresponding KS $p$ values are shown in Appendix~\ref{app:ks_p_values} as a qualitative check. We therefore adopt
$
12.5 < R < 21~\mathrm{kpc}
$
as the fiducial interval for the density fit.

This interval should not be interpreted as proof of absolute completeness. It is the range where the observable RGB luminosity function is most nearly invariant with radius. After the $V_{\max}$ correction, the remaining selection effects are expected to vary slowly enough for estimating $\partial\ln\nu/\partial\ln R$.

\subsubsection{Double exponential fit and density gradient}
\label{sec:dbexpjeans}

\begin{figure}[htb!]
  \centering
  \includegraphics[width=\columnwidth]{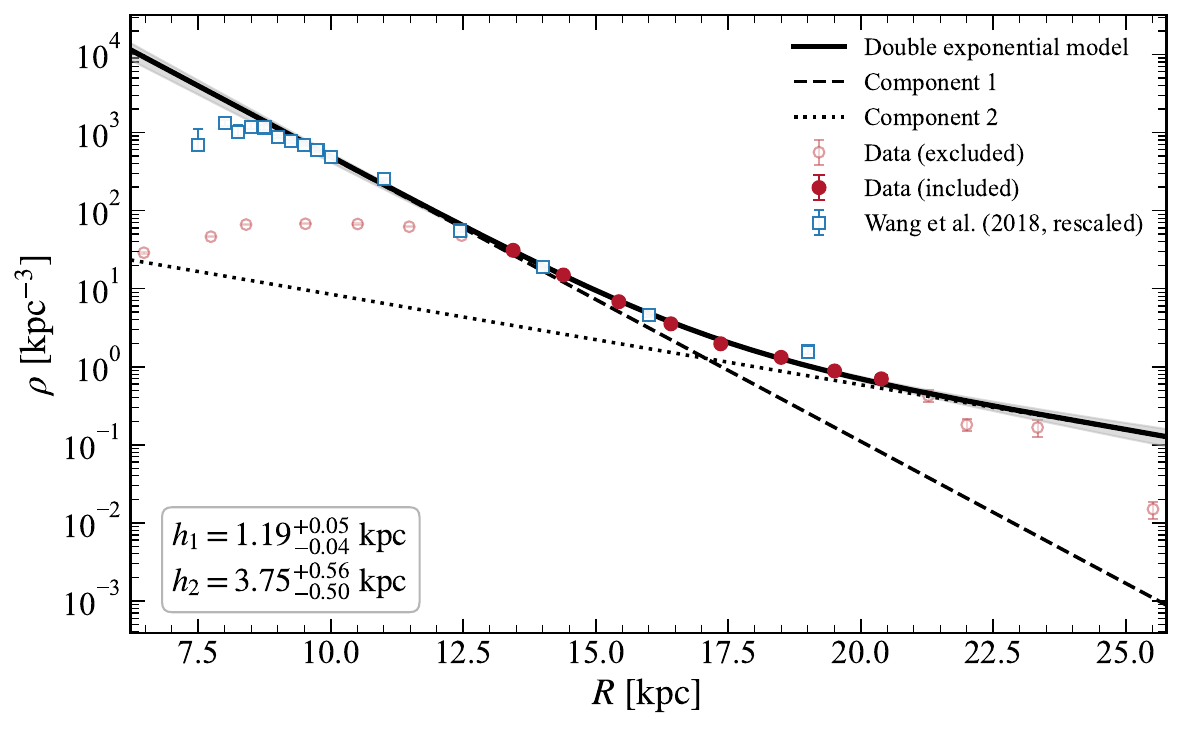}
  \caption{Radial density profile derived from the $V_{\max}$-corrected stellar sample used for the RC measurement. Filled red circles with error bars show the bins included in the fit, while red open circles mark bins excluded because of incompleteness at small and large radii. The solid black line shows the posterior median of the double exponential fit, and the shaded region indicates the $68\%$ credible interval. The dashed and dotted lines show the two exponential components separately. Blue squares show the RGB density profile from \citet{Wang2018}, rescaled by a constant factor for shape comparison.}
  \label{fig:fig_dens}
\end{figure}

Within the validated radial interval, we derive the tracer density profile from the $V_{\max}$ corrected stellar counts in $1\,\mathrm{kpc}$ bins. We assign weighted Poisson uncertainties as described above. The resulting profile is shown in Fig.~\ref{fig:fig_dens}. For comparison, we also show the RGB density profile measured by \citet{Wang2018}, rescaled by a constant factor to match our normalization. This rescaling is allowed because the Jeans equation depends on the logarithmic radial gradient of $\nu(R)$, not on its absolute amplitude. After rescaling, the profile of \citet{Wang2018} agrees well with our measurement over the fitted range. This agreement provides an external check that the radial shape inferred from our $V_{\max}$ corrected APOGEE and \textit{Gaia} sample is consistent with an independent RGB density reconstruction that explicitly modeled survey selection effects. It supports using our corrected profile as an empirical approximation to the intrinsic density gradient of the RGB tracer population over $12.5<R<21\,\mathrm{kpc}$.

\begin{figure}[htb!]
  \centering
  \includegraphics[width=\columnwidth]{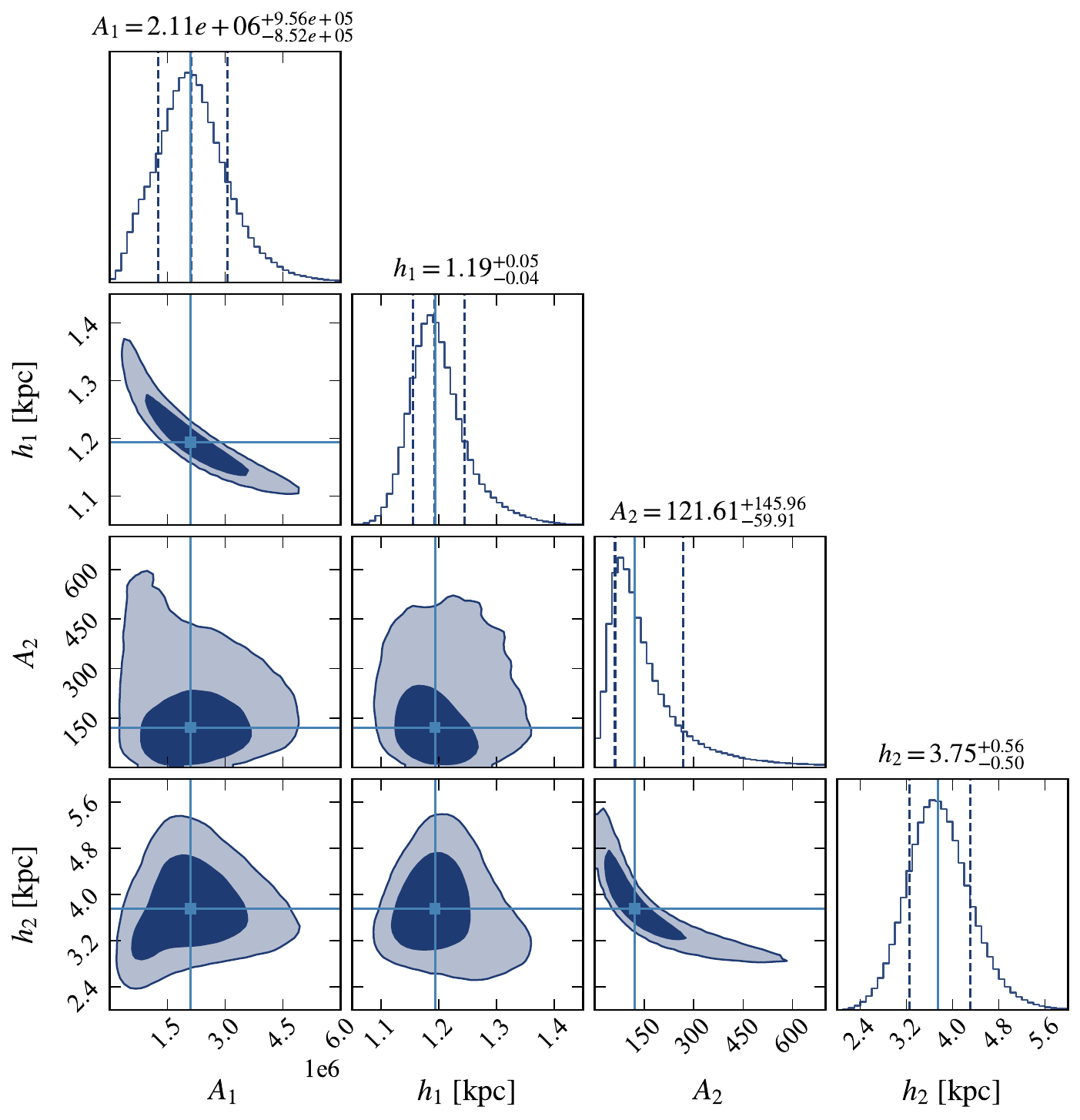}
  \caption{Posterior distributions of the double-exponential fit parameters obtained from the MCMC analysis.}
  \label{fig:fig_dens_corner}
\end{figure}

We model the tracer density as
\begin{equation}
\nu(R)=A_1 \exp\left(-\frac{R}{h_1}\right)+A_2 \exp\left(-\frac{R}{h_2}\right),
\label{eq:double_exp_nu}
\end{equation}
and fit this form with a Markov Chain Monte Carlo (MCMC) procedure using the binned profile and its weighted uncertainties. The posterior distributions are shown in Fig.~\ref{fig:fig_dens_corner}. The two scale lengths are well constrained:
\begin{equation}
h_1 = 1.19^{+0.05}_{-0.04}~\mathrm{kpc}, \qquad
h_2 = 3.75^{+0.56}_{-0.50}~\mathrm{kpc}.
\label{eq:double_exp_scales}
\end{equation}
Over the fitted radial range, the tracer density is better described by a double exponential form than by a single exponential profile.

For the double exponential model in Eq.~(\ref{eq:double_exp_nu}), the logarithmic density gradient entering the Jeans equation is
\begin{equation}
\frac{\partial \ln \nu}{\partial \ln R}
=
-R\,
\frac{
A_1 h_1^{-1}\exp(-R/h_1)+A_2 h_2^{-1}\exp(-R/h_2)
}{
A_1 \exp(-R/h_1)+A_2 \exp(-R/h_2)
}.
\label{eq:dlnnu_double_exp}
\end{equation}
This expression depends only on the shape of the density profile, namely $h_1$, $h_2$, and the relative contribution of the two components through $A_2/A_1$.

Defining
\begin{equation}
\begin{aligned}
w_1(R) &= \frac{A_1\exp(-R/h_1)}{A_1\exp(-R/h_1)+A_2\,\exp(-R/h_2)},
\\
w_2(R) &= \frac{A_2\,\exp(-R/h_2)}{A_1\exp(-R/h_1)+A_2\,\exp(-R/h_2)},
\end{aligned}
\label{eq:weights_double_exp}
\end{equation}

with $w_1+w_2=1$, Eq.~(\ref{eq:dlnnu_double_exp}) becomes
\begin{equation}
\frac{\partial \ln \nu}{\partial \ln R}
=
-R\left(\frac{w_1(R)}{h_1}+\frac{w_2(R)}{h_2}\right).
\label{eq:dlnnu_double_exp_weighted}
\end{equation}
The corresponding circular velocity equation is then
\begin{equation}
\begin{split}
V_{\rm c}^2(R)&= \langle v_\phi^2 \rangle
- \langle v_R^2 \rangle
\left[
1
-
R\left(\frac{w_1(R)}{h_1}+\frac{w_2(R)}{h_2}\right) 
+
\frac{\partial \ln \langle v_R^2 \rangle}{\partial \ln R}
\right] \\
& - R\,\frac{\partial \langle v_R v_z \rangle}{\partial z}.    
\end{split}
\label{eq:jeans_vc_double_exp}
\end{equation}

In the limiting case where one component dominates, Eq.~(\ref{eq:jeans_vc_double_exp}) reduces to the familiar single exponential form.  We adopt this double exponential profile for $\nu(R)$ throughout the fiducial RC calculation. Its density gradient is best constrained over the validated range $12.5<R<21\,\mathrm{kpc}$. Applications outside this interval rely on a smooth extrapolation and are included in the density profile systematic uncertainty.

\subsection{Radial gradient of the radial second moment}
\label{sec:vr2_gradient}

The term $\partial \ln \langle v_R^2 \rangle / \partial \ln R$ in Eq.~(\ref{eq:jeans_vc_general}) describes the radial variation of the radial second velocity moment. Following previous MW Jeans analyses \citep[e.g.][]{Eilers2019,Ou2024}, we model $\sqrt{\langle v_R^2\rangle}$ with a single exponential profile,
\begin{equation}
\sqrt{\langle v_R^2\rangle}
=
\sigma_{R,0}\,\exp\left(-\frac{R}{h_{\sigma}}\right),
\label{eq:sigmaR_exp}
\end{equation}
where $h_{\sigma}$ is the scale length of this profile.

\begin{figure}[htb!]
\centering
\includegraphics[width=\columnwidth]{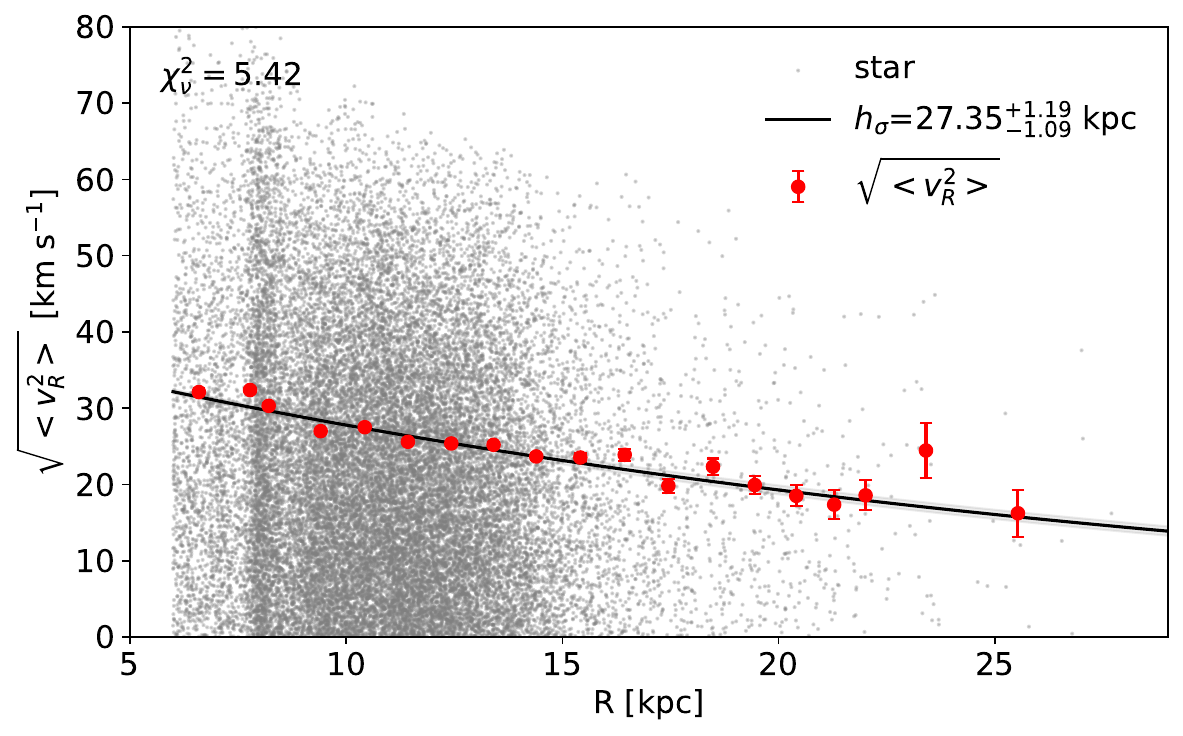}
\caption{Radial profile of $\sqrt{\langle v_R^2\rangle}$ for the stellar sample used in the RC measurement. Gray points show the individual stars, while red points with error bars show the binned estimates. The black line indicates the best fitting single exponential model, and the gray shaded regions show the $1\sigma$ ranges inferred from the MCMC fit.}
\label{fig:vr_r}
\end{figure}

Using the revised radial bins adopted in this work, we recompute $\sqrt{\langle v_R^2\rangle}$ in each bin. The single exponential form still provides an adequate description of the data. Figure~\ref{fig:vr_r} shows the fit and the posterior uncertainty on the scale length. The fit yields $h_{\sigma}=27.35^{+1.19}_{-1.09}\,\mathrm{kpc}$ with reduced $\chi^2_{\nu}=5.42$, indicating that the exponential model captures the overall radial trend but not all local variations, which is accounted for in the systematic uncertainty.

For the model in Eq.~(\ref{eq:sigmaR_exp}), the logarithmic gradient entering the Jeans equation becomes
\begin{equation}
\frac{\partial \ln \langle v_R^2 \rangle}{\partial \ln R}
=
-\frac{2R}{h_{\sigma}}.
\label{eq:dlnvr2}
\end{equation}
Combining this expression with the double exponential tracer density profile from Section~\ref{sec:dbexpjeans}, the circular velocity equation becomes
\begin{equation}
\begin{split}
V_{\rm c}^2(R)&= \langle v_\phi^2 \rangle
- \langle v_R^2 \rangle
\left[
1
-
R\left(\frac{w_1(R)}{h_1}+\frac{w_2(R)}{h_2}\right)
-
\frac{2R}{h_{\sigma}}
\right] \\
&- R\,\frac{\partial \langle v_R v_z \rangle}{\partial z}.
\end{split}
\label{eq:jeans_vc_final}
\end{equation}

The uncertainty on $h_{\sigma}$ is estimated from the MCMC fit and propagated into the statistical uncertainty of the RC through standard error propagation.

\subsection{Cross term in the Jeans analysis}
\label{sec:cross_term_rc}

As discussed in Section~\ref{sec:jeans_framework}, the cross term
$-R\,\partial \langle v_R v_z \rangle / \partial z$ enters Eq.~(\ref{eq:jeans_vc_general}). In many previous MW Jeans analyses, this term was either neglected or treated separately when estimating the systematic uncertainty. Here we include it directly in the fiducial RC. The main practical issue is how to estimate it stably, given the rapidly decreasing number of tracers in the outer disk.

We estimate $\partial \langle v_R v_z \rangle/\partial z$ by dividing the stars in each radial bin into two subsamples above and below the Galactic midplane. We measure $\langle v_R v_z \rangle$ in each subsample and fit a linear relation between $\langle v_R v_z \rangle$ and $z$ with an MCMC procedure. The bin by bin fits are presented in Appendix~\ref{app:cross_term_bin}. Figure~\ref{fig:cross_term_all} shows the resulting radial profile.

\begin{figure}[htb!]
\centering
\includegraphics[width=\columnwidth]{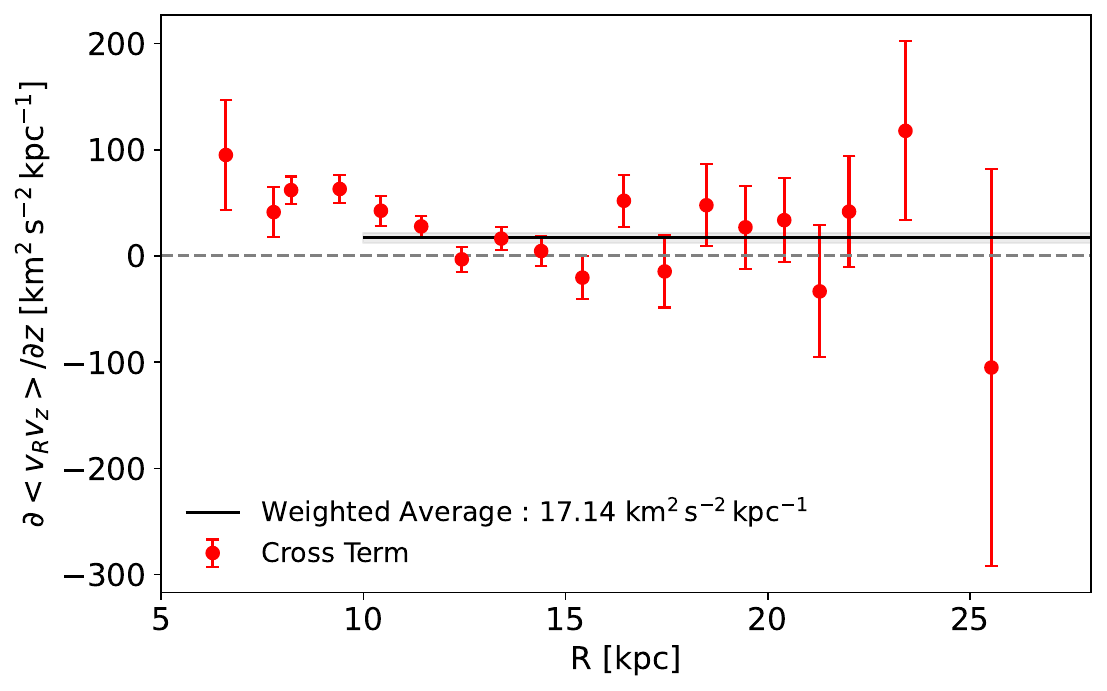}
\caption{Radial profile of $\partial \langle v_R v_z \rangle / \partial z$. Red points show the MCMC estimates in individual radial bins, and the gray dashed line marks zero. The black horizontal line indicates the inverse variance weighted average for the measurements at $R>10\,\mathrm{kpc}$, giving $\langle \partial \langle v_R v_z \rangle / \partial z \rangle_w = 17.14\,\mathrm{km^2\,s^{-2}\,kpc^{-1}}$.}
\label{fig:cross_term_all}
\end{figure}

The measurements show a clear radial variation from $R\simeq6$ to $10\,\mathrm{kpc}$, where the gradient decreases rapidly. This behavior may reflect the influence of the Galactic bar. Beyond $R\simeq10\,\mathrm{kpc}$, the measurements are much flatter on average and remain broadly consistent with a constant value within their uncertainties. At the largest radii, the uncertainties become large because the number of stars per bin is small. Using the local MCMC estimates directly in these outer bins would make the fiducial RC more sensitive to local statistical fluctuations.

We therefore use the local MCMC estimates in the inner disk, where the radial variation is resolved. For $R>10\,\mathrm{kpc}$, we adopt the inverse variance weighted average shown in Fig.~\ref{fig:cross_term_all}. This choice preserves the mean cross term measured in the intermediate and outer disk, while giving little weight to poorly constrained outer bins. The fiducial RC is computed with this prescription.

The uncertainty associated with this prescription is included in the systematic uncertainty. We do not estimate it by replacing the adopted average with the noisy local values in every outer bin. Instead, we use the significance of the local deviations from the adopted average, so that poorly constrained points do not dominate the estimate. The resulting variation is propagated through the Jeans equation and included in Fig.~\ref{fig:systematics_all}. Thus, the cross term is included in the fiducial RC, while the limited precision of the outer disk measurements is carried into the final systematic uncertainty.

\subsection{Systematic uncertainties}
\label{sec:systematics}

\begin{figure}[htb!]
\centering
\includegraphics[width=\columnwidth]{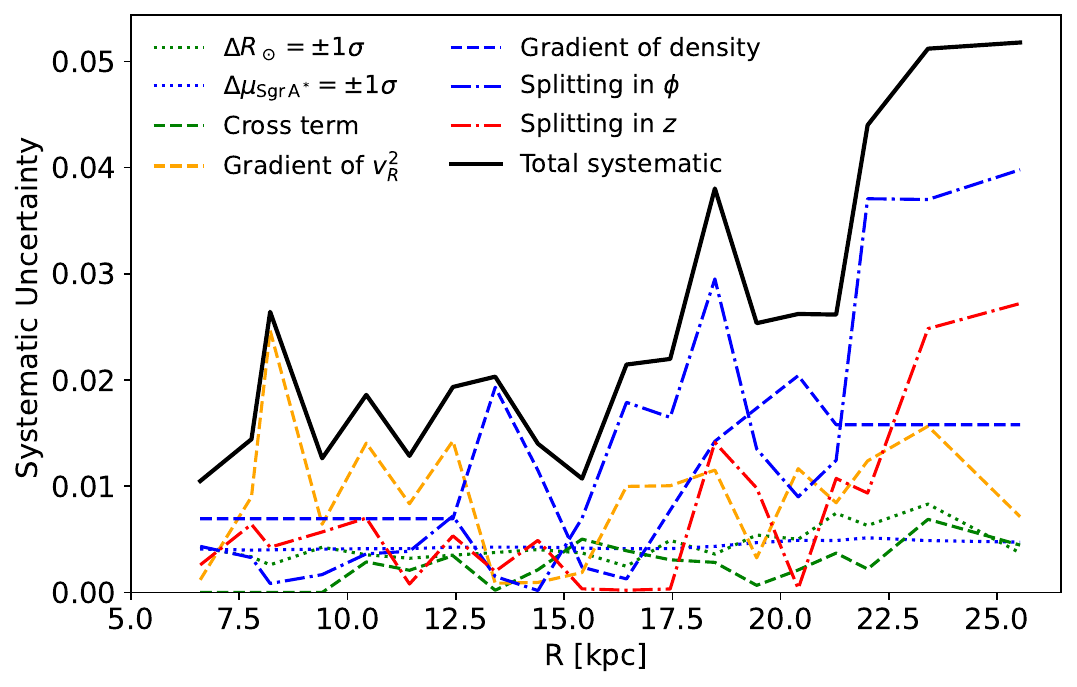}
\caption{Radial dependence of the fractional systematic uncertainties. The dotted lines show the effects of varying $R_\odot$ and the motion of Sgr A$^\star$ by $\pm1\sigma$, following \citet{Eilers2019} and \citet{Ou2024}. The dashed lines show the uncertainties associated with the tracer density gradient, the radial gradient of $\langle v_R^2\rangle$, and the cross term. The dash dotted lines show the effects of splitting the sample by the Galactic azimuth or height. The black solid line shows the total systematic uncertainty, computed as the quadratic sum of all terms.}
\label{fig:systematics_all}
\end{figure}

We define systematic uncertainties as effects that can coherently shift the Jeans-based RC, separate from random measurement errors. The diagonal second moments, $\langle v_\phi^2\rangle$ and $\langle v_R^2\rangle$, are measured directly from the stellar velocities after correcting for the velocity uncertainty covariance, as described in Section~\ref{sec:jeans_framework}. Their measurement uncertainties are already included in the statistical uncertainty of the RC. We therefore do not assign additional systematic terms to these moments themselves. Instead, we consider quantities that affect the velocity transformation, enter the Jeans equation through fitted gradients, or test the spatial dependence of the tracer sample. The resulting fractional contributions are shown in Fig.~\ref{fig:systematics_all}. The total systematic uncertainty is computed as the quadratic sum of the individual terms.

For the Solar parameters, we follow \citet{Eilers2019} and \citet{Ou2024}. We recompute the RC after varying $R_\odot$ and the motion of Sgr A$^\star$ within their quoted $1\sigma$ uncertainties. These changes produce small shifts, typically below the percent level, and remain a minor contribution at all radii.

For the two radial gradient terms, $\partial\ln\nu/\partial\ln R$ and $\partial\ln\langle v_R^2\rangle/\partial\ln R$, we test the sensitivity to the adopted binning and to the use of smooth analytic profiles. We construct an alternative set of $1\,\mathrm{kpc}$ wide bins centred on the weighted mean radii of the fiducial bins, corresponding approximately to a half bin shift. We then estimate local logarithmic slopes from adjacent rebinned measurements with an MCMC linear fit in logarithmic space. The bin by bin measurements and diagnostics are shown in Appendix~\ref{app:gradient_systematics}.

For $\partial\ln\langle v_R^2\rangle/\partial\ln R$, local estimates are available over the full RC range. We therefore recompute the RC after replacing the fiducial exponential gradient with the local rebinned estimate in each radial bin. For the tracer density gradient, the same test is restricted to the validated density fitting range from Section~\ref{sec:density_profile}. Outside this range, the local counts are more affected by selection effects and small number statistics. We therefore extrapolate the density gradient systematic from the fitted region, using representative median fractional RC variations of $\simeq0.007$ for the inner extrapolated region and $\simeq0.015$ for the outer extrapolated region. Compared with earlier analyses that adopted a single exponential tracer profile and varied the scale length by $\pm1\,\mathrm{kpc}$ \citep{Eilers2019,Jiao2023,Ou2024}, this procedure ties the density gradient uncertainty more directly to the measured tracer profile and to its sensitivity to radial binning. As shown in Fig.~\ref{fig:systematics_all}, the density gradient contribution remains moderate at all radii and is smaller than the splitting terms in the outer disk.

The cross term contribution follows the prescription in Section~\ref{sec:cross_term_rc}. Since the cross term is included in the fiducial RC, this systematic term reflects the uncertainty of the adopted prescription rather than the full size of the correction. We estimate it from the significance of the local deviations from the adopted average, so that poorly constrained outer points do not dominate. The resulting contribution is small and varies weakly with radius.

We also perform two splitting tests to estimate the impact of departures from axisymmetry and equilibrium. We split the sample above and below the Galactic plane to test vertical asymmetries, and on the two sides of the Galactic anticentre direction to test azimuthal asymmetries. These tests do not assume a specific perturbation model. They provide an empirical estimate of how much the inferred RC depends on the spatial region sampled by the tracers.

Figure~\ref{fig:systematics_all} shows that the Solar parameters, cross term, and tracer density gradient are not the dominant sources of uncertainty. The total systematic uncertainty is mostly at the percent level inside $R\sim15\,\mathrm{kpc}$, with a local contribution from the $\langle v_R^2\rangle$ gradient near the inner bins. At larger radii, the splitting terms become dominant. The azimuthal split rises beyond $R\sim15\,\mathrm{kpc}$, while the vertical split becomes important beyond $R\sim18$ to $20\,\mathrm{kpc}$. The total systematic uncertainty reaches $\sim4$ to $5\%$ in the outermost bins. Thus, after deriving the tracer density profile from the same stellar sample used for the RC measurement, the main residual systematics are associated with large scale asymmetries in the outer MW disk.

\begin{figure}[htb!]
    \centering
    \includegraphics[width=\columnwidth]{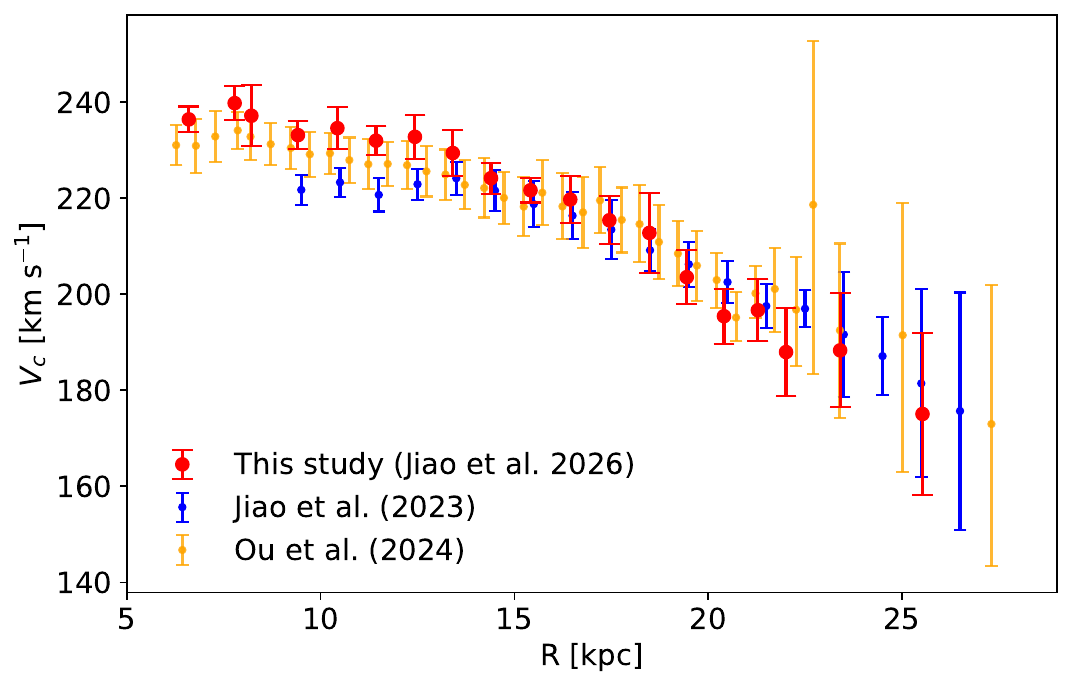}
    \caption{Milky Way rotation curve derived in this work (red circles), compared with \citet{Jiao2023} (blue points) and \citet{Ou2024} (orange points). All error bars include the systematic uncertainty.}
    \label{fig:rc_compare}
\end{figure}

\section{Results}
\label{sec:results}

\subsection{Rotation curve}
\label{sec:results_rc}

Figure~\ref{fig:rc_compare} shows the final MW RC derived in this work. The quoted uncertainties include both the statistical errors and the systematic uncertainties discussed in Section~\ref{sec:systematics}. The measured RC is listed in Table~\ref{tab:rc_values}. It extends from $R=6.60$ to $25.53\,\mathrm{kpc}$, with $V_{\rm c}$ decreasing from $236.43\pm2.69\,\mathrm{km\,s^{-1}}$ in the innermost bin to $175.07\pm16.86\,\mathrm{km\,s^{-1}}$ in the outermost bin. The curve is nearly flat only in the inner disk and then declines with radius. The decline becomes clear beyond $R\sim14$ to $15\,\mathrm{kpc}$.

Compared with \citet{Ou2024}, our RC is higher in the inner disk. This difference is expected from the tracer density term adopted in the Jeans analysis. Instead of assuming a fixed single exponential profile, we derived the density profile from the selected RGB sample within the range $12.5< R < 21$ kpc (see Fig.~\ref{fig:fig_dens}) and found that it is better described by a double exponential form, with a short inner scale length of $h_1=1.19^{+0.05}_{-0.04}\,\mathrm{kpc}$ and a larger outer component of $h_2=3.75^{+0.56}_{-0.50}\,\mathrm{kpc}$. The short inner scale length gives a steeper radial density gradient in the Jeans equation, which leads to a higher inferred $V_{\rm c}$ in the inner Galaxy than in analyses adopting a shallower fixed exponential profile. The higher inner RC obtained here is therefore a consequence of using a tracer density profile constrained by the same data.

At larger radii, our RC follows the same declining trend as \citet{Jiao2023} and \citet{Ou2024}. This agreement is clearest beyond $R\sim19\,\mathrm{kpc}$, where the three curves decrease with radius within their uncertainties. For example, our RC decreases from $203.57\,\mathrm{km\,s^{-1}}$ at $R=19.45\,\mathrm{kpc}$ to $195.44\,\mathrm{km\,s^{-1}}$ at $R=20.41\,\mathrm{kpc}$ and to $175.07\,\mathrm{km\,s^{-1}}$ at $R=25.53\,\mathrm{kpc}$. The outer decline reported in previous \textit{Gaia} DR3 studies is therefore retained after the revised tracer density profile, wider radial binning, and updated systematic uncertainty. We quantify the slope of this decline in Section~\ref{sec:keplerian_decline}.

The comparison in Fig.~\ref{fig:rc_compare} should be interpreted with the different radial sampling in mind. Our outer disk bins are broader than those of \citet{Ou2024}, because they are chosen to better match the radial precision of the data and to keep enough tracers per bin. The points therefore should not be compared one by one. The relevant result is that the global outer disk behavior is consistent among the three RC measurements.

The revised tracer density profile mainly affects the inner RC, where the Jeans density term is more sensitive to the adopted radial profile. In contrast, the declining outer disk trend remains after the updated density model and propagated systematic uncertainties are included.

\begin{deluxetable}{ccc}
\tablecaption{Final MW RC measurements\label{tab:rc_values}}
\tablehead{
\colhead{$R$} & \colhead{$V_{\rm c}$} & \colhead{$e_{V_{\rm c}}$\tablenotemark{a}} \\
\colhead{(kpc)} & \colhead{($\mathrm{km\,s^{-1}}$)} & \colhead{($\mathrm{km\,s^{-1}}$)}
}
\startdata
6.60  & 236.43 & 2.69  \\
7.78  & 239.82 & 3.53  \\
8.21  & 237.20 & 6.31  \\
9.41  & 233.16 & 2.99  \\
10.43 & 234.61 & 4.39  \\
11.43 & 232.00 & 3.01  \\
12.43 & 232.80 & 4.54  \\
13.41 & 229.43 & 4.70  \\
14.40 & 224.20 & 3.22  \\
15.42 & 221.68 & 2.56  \\
16.44 & 219.74 & 4.87  \\
17.45 & 215.42 & 5.03  \\
18.48 & 212.79 & 8.38  \\
19.45 & 203.57 & 5.66  \\
20.41 & 195.44 & 5.75  \\
21.28 & 196.68 & 6.49  \\
22.01 & 187.96 & 9.18  \\
23.41 & 188.32 & 11.89 \\
25.53 & 175.07 & 16.86 \\
\enddata
\tablenotetext{a}{The uncertainty includes the statistical uncertainty and the total systematic uncertainty adopted in this work.}
\end{deluxetable}

\subsection{Mass models}
\label{sec:mass_models}

To infer the MW mass distribution from the RC derived in this work, we adopt a parametric model with baryonic and DM components. For the baryons, we use the B2 model of \citet{deSalas2019}, which provides an axisymmetric description of the stellar, gaseous, and dust components of the Galaxy.

The bulge is described by a Hernquist potential,
\begin{equation}
\Phi_{\rm bulge}(r) = -\frac{G M_{\rm bulge}}{r+r_{\rm b}},
\end{equation}
with $M_{\rm bulge}=1.55\times10^{10}\,M_{\odot}$ and $r_{\rm b}=0.7\,\mathrm{kpc}$. The disk components are described by a double-exponential density profile,
\begin{equation}
    \rho (R,z) = \rho_0 \exp \left( -\frac{R}{R_{\rm d}} - \frac{|z|}{z_{\rm d}} \right),
    \label{eq:doubleexponential}
\end{equation}
where $\rho_0 = M/(4\pi z_{\rm d} R_{\rm d}^2)$, $M$ is the component mass, and $R_{\rm d}$ and $z_{\rm d}$ are the radial scale length and vertical scale height, respectively. The adopted B2 parameters are listed in Table~\ref{tab:baryon_b2}. The cold and warm dust components have negligible effect on the RC, but are included in the total baryonic mass for consistency. The total baryonic mass of the adopted B2 model is $M_{\rm bar}=6.16\times10^{10}\,M_\odot$.

\begin{deluxetable}{lccc}
\tablecaption{Disk components of the adopted B2 baryonic model\label{tab:baryon_b2}}
\tablehead{
\colhead{Component} & \colhead{$M$} & \colhead{$R_{\rm d}$} & \colhead{$z_{\rm d}$} \\
\colhead{} & \colhead{($M_\odot$)} & \colhead{(kpc)} & \colhead{(kpc)} \\
}
\startdata
Stellar disk        & $3.65\times10^{10}$ & 2.35  & 0.14 \\
$\mathrm{H\,I}$ gas & $8.2\times10^{9}$   & 18.24 & 0.52 \\
$\mathrm{H}_2$ gas  & $1.3\times10^{9}$   & 2.57  & 0.08 \\
Cold dust           & $7.0\times10^{7}$   & 5.00  & 0.10 \\
Warm dust           & $2.2\times10^{5}$   & 3.30  & 0.09 \\
\enddata
\end{deluxetable}

For the DM halo, we adopt an Einasto profile \citep{Einasto1965,Retana-Montenegro2012} as the fiducial model,
\begin{equation}
    \rho(r) = \rho_0 \exp\left[-\left(\frac{r}{h}\right)^{1/n}\right],
    \label{eq:einasto}
\end{equation}
where $h$ is the scale radius and $n$ is the Einasto index. Previous MW RC analyses have found that Einasto like profiles can fit the MW circular velocity curve better than the Navarro--Frenk--White profile \citep[NFW;][]{Navarro1997,Chemin2011,Jiao2021}. For comparison, we also fit an NFW halo,
\begin{equation}
\rho_{\rm NFW}(r)=\frac{\rho_0}{(r/r_{\rm s})(1+r/r_{\rm s})^2}.
\end{equation}

We fit the halo parameters with the affine invariant MCMC sampler \textsc{ emcee}\footnote{\url{https://github.com/dfm/emcee}} \citep{emcee2013}. For the Einasto model, we use flat priors on $M_0=4\pi h^3\rho_0$, $h$, and $1/n$, with $10^{10}<M_0/M_\odot<10^{14}$, $0<h<20\,\mathrm{kpc}$, and $0<1/n<5$. The likelihood is
\begin{equation}
    \ln{\mathcal{L}} = -\frac{1}{2} \sum_i \left( \frac{v_{\mathrm{mod},i}-v_{\mathrm{obs},i}}{\sigma_i} \right)^2,
    \label{eq:likelihood}
\end{equation}
where $v_{\mathrm{mod},i}$ is the model circular velocity, $v_{\mathrm{obs},i}$ is the measured RC, and $\sigma_i$ is the adopted uncertainty of the $i$th RC point.

We extrapolate the DM halo to the virial radius, denoted $R_{\rm vir}\equiv R_{200}$. This radius is defined such that the mean DM density enclosed within it is 200 times the critical density of the Universe. We adopt $\rho_{\rm cr}=1.34\times10^{-7}\,M_\odot\,\mathrm{pc}^{-3}$ \citep{hinshaw2013}. We denote the DM mass inside $R_{\rm vir}$ as $M_{\rm vir}$, and the total dynamical mass ($M_{\rm dyn}$, or simply the total mass $M_{\rm tot}$) as the sum of the baryonic mass and the virial DM mass.

\begin{figure}[htb!]
    \centering
    \includegraphics[width=\columnwidth]{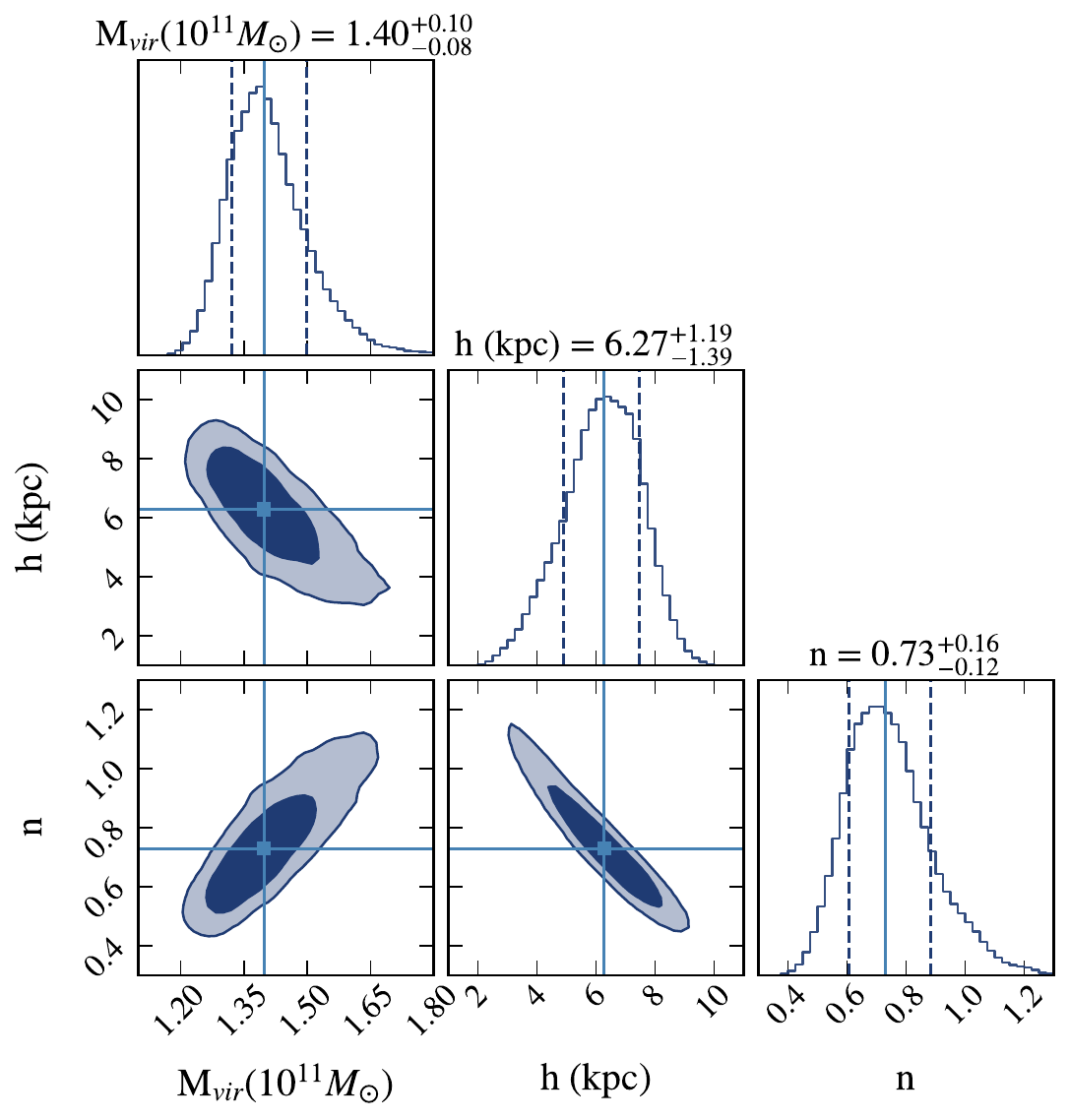}
    \caption{Posterior distributions of the Einasto halo parameters fitted to the RC derived in this work.}
\label{fig:mc_dis_eina}
\end{figure}

\begin{figure}[htb!]
    \centering
    \includegraphics[width=\columnwidth]{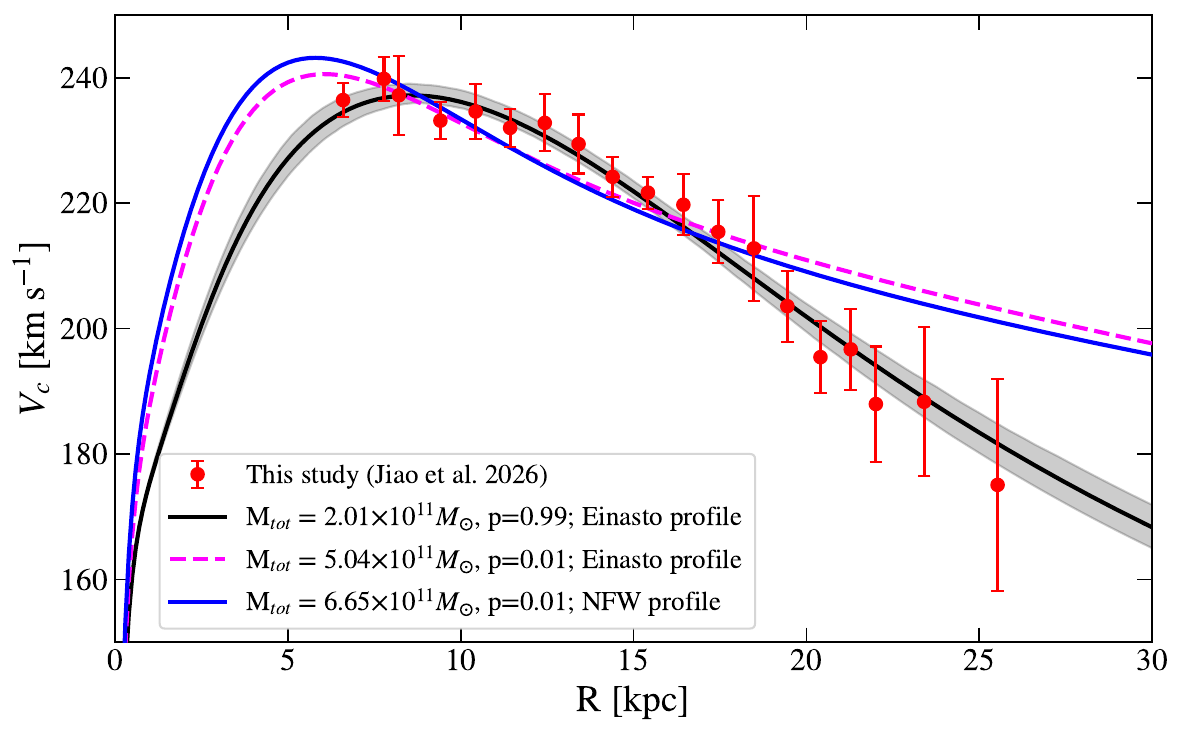}
    \caption{Mass model fits to the RC derived in this work. The black line shows the best fitting Einasto model, with a $\chi^2$ probability of $p=0.99$. The blue dashed line shows the highest mass Einasto model allowed at $p=0.01$. The best-fitting NFW model is also shown as a solid blue line for comparison.}
\label{fig:eina_fit}
\end{figure}

Figure~\ref{fig:mc_dis_eina} shows the posterior distributions of the Einasto halo parameters. The mass shown in this posterior is the DM virial mass. After adding the B2 baryonic mass, we obtain
$
M_{\rm dyn}=2.01^{+0.10}_{-0.08}\times10^{11}\,M_\odot.
$ The best-fitting Einasto and NFW models are shown in Fig.~\ref{fig:eina_fit}. The NFW profile reaches a higher mass but provides a poorer fit to the outer RC.

We also estimate the maximum mass allowed by the RC for a given fit quality. We use the $\chi^2$ probability,
\begin{equation}
P(\chi^2, \nu) = \int_{\chi^2}^{\infty} \frac{1}{2^{\nu/2} \, \Gamma(\nu/2)} \, x^{\nu/2 - 1} e^{-x/2} \, dx,
\end{equation}
with
\begin{equation}
\chi^2 = \sum_i \frac{(v_{\mathrm{obs},i}-v_{\mathrm{mod},i})^2}{\sigma_i^2},
\end{equation}
where $\nu$ is the number of degrees of freedom. We then use simulated annealing \citep{Kirkpatrick1983} to find the Einasto parameters that maximize $M_{\rm dyn}$ while satisfying a given probability threshold. For a minimum fitting probability of $p=0.01$, we find $M_{\rm max}(p=0.01)=5.04\times10^{11}\,M_\odot$.
The corresponding RC is also shown in Fig.~\ref{fig:eina_fit}.

\subsection{Keplerian decline}
\label{sec:keplerian_decline}

The RC derived in this work preserves the outer decline found in recent \textit{Gaia} DR3 studies \citep{Jiao2023,Ou2024}. To quantify its slope, we fit the outer RC with a power law,
\begin{equation}
V_{\rm c}(R)=A R^\gamma ,
\label{eq:powerlaw_vc}
\end{equation}
where $\gamma=-1/2$ corresponds to a Keplerian decline and $\gamma=0$ to a flat RC.

We tested several inner boundaries for the outer power law fit. As a fiducial choice, we adopt $R\simeq16\,\mathrm{kpc}$, approximately where the RC enters a monotonic outer decline. We regard this choice as an empirical definition of the fitted range, not a sharp physical transition radius.

\begin{figure}[htb!]
\centering
\includegraphics[width=0.85\columnwidth]{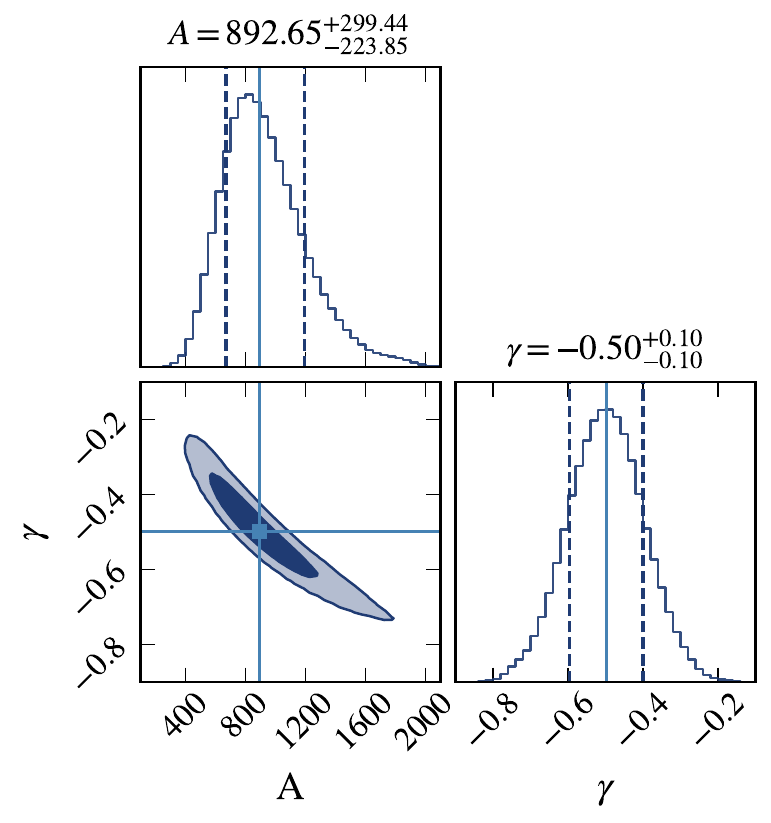}
\caption{Posterior distributions of the power law parameters $A$ and $\gamma$ for the outer RC fit in Eq.~(\ref{eq:powerlaw_vc}). The solid lines mark the posterior medians, and the dashed lines indicate the $1\sigma$ ranges. The inferred slope is consistent with the Keplerian value $\gamma=-1/2$.}
\label{fig:kep_mcmc}
\end{figure}

The posterior distributions of $A$ and $\gamma$ are shown in Fig.~\ref{fig:kep_mcmc}. We find
$
\gamma=-0.50^{+0.10}_{-0.10}
$
consistent with a Keplerian decline over the adopted fitting interval. A flat RC is strongly disfavored at the 5$\sigma$ level over the same interval.

\begin{figure}[htb!]
\centering
\includegraphics[width=\columnwidth]{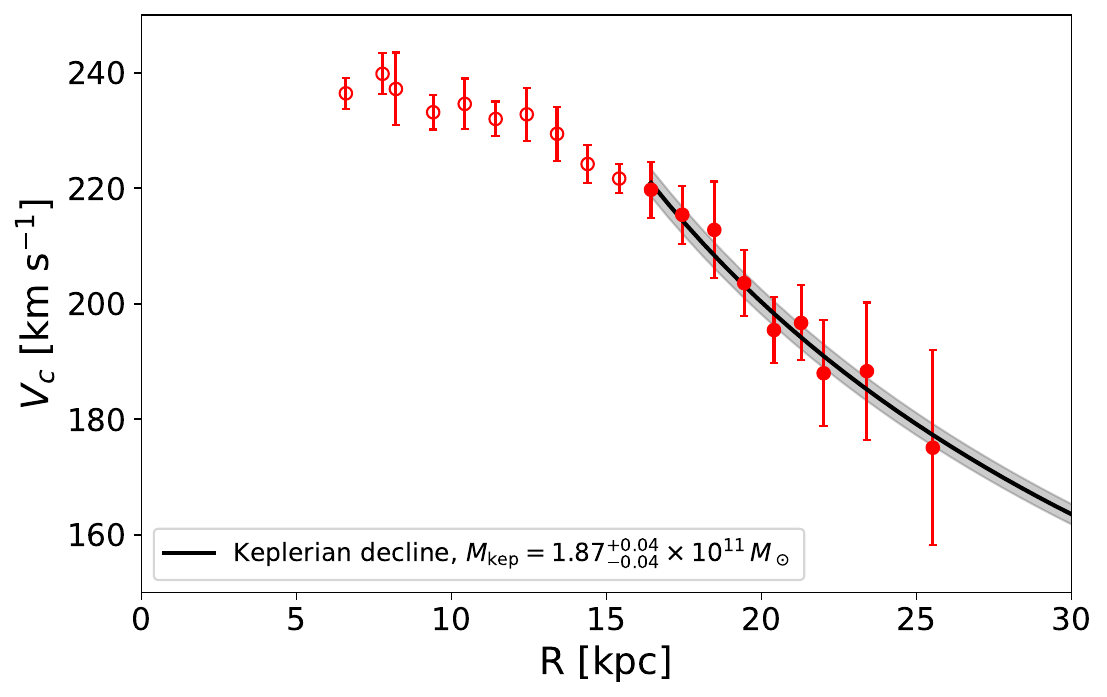}
\caption{Fixed Keplerian fit to the outer part of the MW RC derived in this work. Open red circles show RC points not included in the fit, while filled red circles show the outer points used for the fit. The black curve shows the best--fitting Keplerian decline, and the gray band shows the $1\sigma$ uncertainty.}
\label{fig:kep_fit}
\end{figure}

For comparison, we also fit the same outer interval with a fixed Keplerian form. The best fitting curve is shown in Fig.~\ref{fig:kep_fit}. A fixed Keplerian form over the adopted outer interval gives an effective enclosed mass of
$
M_{\rm kep}=1.87^{+0.04}_{-0.04}\times10^{11}\,M_\odot .
$
This value is close to the dynamical mass inferred from the fiducial Einasto model in Section~\ref{sec:mass_models}. This quantity should be interpreted only as a local description of the slowly increasing enclosed mass over the disk range, not as a model-independent virial mass.

\section{Discussion}
\label{sec:discussion}

\subsection{Is the Milky Way disk truncated?}
\label{sec:discussion_truncation}

The possibility of a truncated outer MW disk must be considered, because the tracer density enters the radial Jeans equation through the logarithmic slope $\partial \ln \nu / \partial \ln R$. If the true tracer density steepens sharply at large radii while the adopted model is too shallow, the inferred $V_{\rm c}(R)$ can be biased low. This is the main concern raised by \citet{Koop2024}. The relevant question here is therefore whether the RGB tracer population used in this work shows evidence for such a sharp outer cutoff.

Outer disks of spiral galaxies are often more complex than either a single exponential profile or an abrupt truncation. \citet{PohlenTrujillo2006} found that only $\sim10\%$ of nearby late type spirals are consistent with a pure exponential profile, usually referred to as Type~I. Most systems instead show broken profiles, with Type~II down bending profiles and Type~III up bending profiles accounting for $\sim60\%$ and $\sim30\%$ of their sample, respectively. Such breaks do not necessarily imply a sudden loss of stellar mass. \citet{Bakos2008} showed that breaks in surface brightness profiles can be much weaker in stellar mass surface density profiles. For Type~II galaxies, they found a stellar mass surface density of $13.6\pm1.6\,M_\odot\,\mathrm{pc}^{-2}$ at the break and $14.7\pm1.2\%$ of the stellar mass beyond the break radius. This distinction matters for Jeans modeling, because the correction depends on the tracer density rather than on the light profile alone.

Recent MW measurements point in the same direction. Using LAMOST RGB stars, \citet{Wang2018} found that the Galactic disk extends to at least $\sim19$~kpc and shows two breaks near $R\sim11$ and $\sim14$~kpc. The corresponding radial scale lengths are $2.12\pm0.26$~kpc, $1.18\pm0.08$~kpc, and $2.72$~kpc in the three radial intervals. More recently, \citet{Lian2024} also found that the MW surface brightness profile is better described by a broken exponential than by a single exponential, with a half light radius of $5.75\pm0.38$~kpc. These results favor a structured outer disk rather than a simple sharp truncation.

Our tracer density fit gives a consistent conclusion. We measured $\nu(R)$ directly from the same low $\alpha$ upper RGB stars used in the Jeans analysis. Over $12.5<R<21$~kpc, where the sample is least affected by incompleteness, the density profile is well described by a double exponential with $h_1=1.19^{+0.05}_{-0.04}\,{\rm kpc}$ and $h_2=3.75^{+0.56}_{-0.50}\,{\rm kpc}.$ The shorter scale length agrees closely with the intermediate segment measured by \citet{Wang2018}, $1.18\pm0.08$~kpc. The longer component is more extended than their outer value of $2.72$~kpc, but it supports the same qualitative picture: the outer MW disk is not well described by a single exponential, yet it remains extended rather than sharply truncated. In the radial range where our density measurement overlaps the intermediate and outer segments of \citet{Wang2018}, the star counts are reproduced by a smooth double exponential profile.

The measured density profile addresses the main concern raised by \citet{Koop2024}, namely that an unmodeled tracer density truncation could make a Jeans-based RC appear artificially declining. In our case, the measured tracer profile does not show such a sharp cutoff over the range where the density can be constrained most reliably. Once this density profile is propagated through the Jeans equation, the density related uncertainty remains modest and is smaller than the splitting systematics in the outer disk. An unmodeled sharp truncation is therefore unlikely to be the dominant origin of the declining RC found here.

This does not imply that the outer MW disk is perfectly axisymmetric or in steady state. The outer disk is warped, flared, and asymmetric \citep{Bland-Hawthorn2016,Antoja2021}, and these effects can affect Jeans-based RC measurements \citep{Koop2024}. Our conclusion is more limited. Within the radial range where the tracer density can be measured most reliably, the data favor a structured and extended disk with a smoothly varying density gradient. A further break beyond $R\gtrsim21\,\mathrm{kpc}$ cannot be excluded. However, the present density measurements do not support a sharp tracer truncation as the main explanation for the observed outer RC decline.

\subsection{Equilibrium condition of the Milky Way disk}
\label{sec:discussion_equilibrium}

A key assumption of a Jeans Analysis of the MW RC is that the disk can be approximated as an axisymmetric system close to dynamical equilibrium. However, the hierarchical scenario predicts that the MW has experienced several encounters that may have affected its disk equilibrium. Its last major merger, associated with GSE, occurred roughly $9$ to $10$~Gyr ago \citep{Belokurov2018,Helmi2018,Haywood2018}. This is much longer than the orbital time of the present disk. Even at $R\simeq30$~kpc, stars have had time to complete about $5$ to $6$ orbital cycles, suggesting that they have had sufficient time to relax toward virial equilibrium \citep{Gnedin1999,Hammer2025}.

The passages of the Sagittarius (Sgr) dwarf galaxy through the MW disk can create such a perturbation. \citet{Laporte2018} proposed that Sgr-like encounters can excite vertical motions of order $\sim10\,\mathrm{km\,s^{-1}}$ near the solar neighbourhood, with larger amplitudes in the far outer disk. \citet{Koop2024} further argued that departures from equilibrium and axisymmetry can bias Jeans-based estimates of the circular velocity by $\sim10\%$ to $15\%$ for the MW. However, \citet{Hammer2024Simu} verified that some simulations may overestimate satellite perturbations if their subhalo populations are dynamically hotter and more massive than the observed MW satellites. Some perturbation models adopt an initial Sgr mass of several $10^{10}\,M_\odot$ \citep{Laporte2018}, while other reconstructions favor much lower values \citep{WangHF2022,Vasiliev2021}. Thus, Sgr probably contributes to the observed outer disk asymmetries, but the amplitude of its effect on the Jeans derived RC remains uncertain.

The Magellanic Clouds (MCs) may also contribute to the large-scale response of the MW disk to their passage as proposed by \citet{Conroy2021}. MCs are likely recent infallers and have very recently passed pericentre \citep{Wang2019,Vasiliev2023}, and their influence on the disk kinematics may therefore be limited. In this sense, the MW disk may be relatively less affected by strong recent tidal perturbations than M31, where a recent merger model produces non-equilibrium and oscillatory gas motions at the disk outskirts \citep{Hammer2025}. 

\textit{Gaia} observations have scrutinized the detailed properties of the 6D MW disk, including phase space spirals, ridges, arches, and large asymmetries along various azimuthal directions, and especially between regions above and below the disk plane \citep{Antoja2018,Hunt2019,Katz2018,Antoja2021,Drimmel2023}. These observations suggest that the disk is perturbed, but not dominated by these perturbations over most of the radial range used here. Toward the Galactic anticentre, \citet{Antoja2021} found north--south asymmetries in the median vertical and rotational velocities, including differences of order several to $10\,\mathrm{km\,s^{-1}}$. They also identified an outer disk component, mainly below the plane, with a rotation velocity offset of order $30\,\mathrm{km\,s^{-1}}$ and positive vertical motion. Using \textit{Gaia} DR3, \citet{Drimmel2023} also found that, beyond $R\sim11\,\mathrm{kpc}$ toward the anticentre, stars below the plane tend to rotate faster than those above it. These results show that the outer disk is not in perfect equilibrium, and it remains unclear which merger event, if any, produced each perturbation. Simulations of dwarf passages near the disk predict bending, breathing, and warp, but the amplitudes of these effects are model-dependent.

We tested whether similar structures are present in our selected tracer sample. In Appendix~\ref{app:vphi_r_diagnostics}, we show the distribution in the $R$ and $-v_\phi$ plane, with bins coloured by height, vertical velocity, vertical velocity dispersion, and number of stars. We perform this test both for a near plane subsample, $|Z|<0.25\,\mathrm{kpc}$, and for the full sample. In contrast to the strong structures reported in the much broader \textit{Gaia} samples, our selected tracers do not show a clear branch in which stars below the plane rotate systematically faster than those above it. The main visible departure is a weak bimodal structure in the $v_z$ map. This indicates that the disk is not perfectly phase mixed, but it also shows that our tracer selection reduces the strongest non-equilibrium features in the kinematic sample used for the RC.

The selected sample should therefore be regarded as suitable for an azimuthally averaged Jeans estimate in an approximate sense, with residual asymmetries propagated into the systematic uncertainty estimate. We address this in two ways. First, we select tracers that are best suited to an axisymmetric Jeans analysis. \citet[][their Appendix A]{Ou2024} showed that beyond $R=20\,\mathrm{kpc}$ nearly all stars in their disk sample follow nearly circular orbits, with only 3 out of 244 outer disk stars having highly eccentric orbits. Our tangential velocity cut and outlier clipping further remove counter rotating and strongly non circular contaminants. Second, we use splitting tests to estimate how much the inferred RC depends on the sampled region of the disk. We split the sample above and below the Galactic plane to test vertical asymmetries, and on the two sides of the Galactic anticentre direction to test azimuthal asymmetries.

The systematics shown in Fig.~\ref{fig:systematics_all} reflect these effects. The splitting terms are small in the inner disk but increase at large radii, where they dominate the systematic uncertainty. This behavior is expected if the MW disk is close to equilibrium in an average sense over the inner and intermediate disk, while becoming more easily perturbed in the outer regions. As an additional check, Appendix~\ref{app:systematic_z} shows that a stricter vertical selection, $|z|<1\,\mathrm{kpc}$, gives an RC consistent with the fiducial $|z|<2\,\mathrm{kpc}$ result over their common radial range. This supports that the outer decline is not driven by the adopted vertical cut, although the residual vertical asymmetry is still included through the splitting tests.

We therefore do not assume that the MW disk is perfectly stationary or perfectly axisymmetric, especially beyond $R\sim10$ to $12$~kpc and toward the anticentre. Instead, we account for this limitation through tracer selection, a density profile measured from the same sample, and vertical and azimuthal splitting tests. With these choices, the Jeans formalism remains a useful tool for measuring the large scale MW RC. The increasing systematic uncertainty in the outermost bins reflects the decreasing accuracy of the equilibrium approximation, rather than removing the information carried by the disk kinematics.


\subsection{Milky Way mass comparison with other studies}
\label{sec:mw_mass_comparison}

Our best fit Einasto model gives a low MW mass, $M_\mathrm{tot}= 2.01^{+0.10}_{-0.08} \times10^{11} M_\odot$. This result is consistent with previous analyses that found a declining outer MW RC \citep{Jiao2021,Jiao2023,Wang2023,Ou2024}. In particular, the Keplerian decline beyond $R\sim16$ kpc implies that the enclosed mass increases only slowly over the radial range directly probed by the disk tracers (See Fig.~\ref{fig:kep_fit}). Even if we consider a very poorly fit model with a fitting probability of 0.01, the maximum allowed mass is 5.04 $\times10^{11} M_\odot$ (See Fig.~\ref{fig:eina_fit}).

The disk RC directly constrains the gravitational field only out to the last measured point, at $R\sim25$ to $26$ kpc. By contrast, $M_{200}$ depends on the halo profile out to radii of order $\sim100$ to $200$ kpc. The inferred virial mass is therefore model-dependent, particularly through the assumed outskirt slope of the DM profile \citep{Jiao2021}, as we find in the present work. When the halo is forced to follow an NFW profile, the best fit mass rises to $M_{200}=6.65\times10^{11}\,M_\odot$, but the fit is poor, with $p=0.01$ (see Fig.~\ref{fig:eina_fit}). Thus, for a declining outer RC, NFW or gNFW extrapolations have to provide higher masses than Einasto like profiles, without fitting well the outermost RC measurements.

This difference also appears in previous RC studies. \citet{Sylos_Labini2023} fitted the \textit{Gaia} DR3 RC with an NFW halo and obtained $M_{\rm vir}=(6.5\pm0.3)\times10^{11}\,M_\odot$. \citet{Cautun2020} adopted a contracted halo model and inferred $M_{200}^{\rm tot}=1.08^{+0.20}_{-0.14}\times10^{12}\,M_\odot$. In contrast, \citet{Ou2024} found that a cored Einasto profile fits the MW circular velocity curve better than a generalized NFW profile, giving a virial DM mass of $1.81^{+0.06}_{-0.05}\times10^{11}\,M_\odot$. Our result agrees with this low mass Einasto solution, but uses an independent treatment of the tracer density term and of the main outer disk systematics.

The \textit{Gaia} DR3 escape velocity analysis of \citet{Roche2024} inferred $M_{200c}\approx0.64\times10^{12}\,M_\odot$ under an NFW assumption, comparable to our result for the same profile. Globular cluster (GC) kinematics provide an independent comparison but remain sensitive to tracer selection and the adopted Galactic potential. Excluding Crater and Pyxis, \citet{Wang2022} reduced their Einasto estimate from $5.73^{+0.76}_{-0.58}$ to $5.36^{+0.81}_{-0.68}\times10^{11}\,M_\odot$, while \citet{Hammer2024} found the GC eccentricity distribution to be compatible with MW masses of $2-16\times10^{11}\,M_\odot$ after excluding these two outer GCs, Crater and Pyxis, which may be on their first infall \citep[][and references therein]{Hammer2024}. Halo star analyses typically find $M(<100\,{\rm kpc})\sim(6$ to $7)\times10^{11}\,M_\odot$ and virial masses of order $\sim10^{12}\,M_\odot$ \citep{Deason2021,Bird2022,Shen2022}. Assuming that dwarf galaxies are in equilibrium systematically leads to very high MW masses, for example $M=1.23^{+0.21}_{-0.18}\times10^{12}\,M_\odot$ from \citet{li2020} and $M_{\rm vir,\Delta=97}=1.51^{+0.45}_{-0.40}\times10^{12}\,M_\odot$ from \citet{Fritz2020}. On one hand, these methods probe larger radii than disk stars and may appear to be more sensitive to the total halo mass. On the other hand, estimates based on dwarf orbits can be affected by the tracer equilibrium state, while  halo stars at the very outskirts are known to be affected by the Large Magellanic Cloud \citep[LMC,][]{Chandra2023,Han2025}. 

The equilibrium assumption is especially important for dwarf galaxy tracers. 
\citet{Gott1975} introduced the idea that later-accreted material is less tightly bound, which was subsequently extended into a tight relationship between the orbital energy and the infall epoch of satellite galaxies \citep[e.g.,][]{Rocha2012}. \textit{Gaia} now allows this test, and the MW accretion history has been established by \citet{Hammer2023,Hammer2024Simu,Hammer2024} from ancient events such as Kraken-lowE and GSE to Sgr and LMC infall. Dwarf orbital energies are found to be 5 times (3 times) larger than those of the GSE (Sgr infall) events, respectively. This suggests a late arrival, less than 3 Gyr ago, for dwarf galaxies, similar to that of the LMC.
In such a case, dwarf galaxy orbits are not necessarily virialized tracers of the MW potential, because they did not have time to trace a single orbit. If they are treated as an equilibrium satellite population, satellite based mass estimates may be biased high. 

The difference between the low mass inferred from the disk RC and the higher masses inferred from halo or satellite tracers reflects the different radial ranges and dynamical assumptions of these methods. Our result is based on the disk kinematics and shows that the MW mass profile rises only slowly over the measured RC range. A halo mass close to $10^{12}\,M_\odot$ would require substantial additional mass beyond the last disk point while producing only a weak signature over the measured outer RC range. More complex outer halo structures, such as shells or rings, could in principle increase the mass at large radii \citep[e.g.,][]{Huang2016}. They are not required by the present data and may be difficult to reconcile with a long lived, quasi-equilibrium disk.

Another estimate of the MW mass is obtained by combining the mass fraction of M31 and the MW with the total Local Group (LG) mass inferred from the Hubble--Lema\^itre flow \citep{Penarrubia2014, Makarov2025}, LG momentum balance \citep{Diaz2014}, and the Timing Argument \citep{vanDerMarel2012, Benisty2024}. These methods generally favor a relatively high MW mass ($\sim10^{12}\,M_\odot$) if all of the LG mass is primarily assigned to MW and M31. However, these LG mass estimates remain consistent with the lower MW and M31 masses inferred from rotation curves, once the significant contribution from the surrounding circumgalactic medium (CGM) and intergalactic medium (IGM) is taken into account \citep{Akib2025}.

We therefore interpret the declining MW RC as favoring a low mass extrapolation when an Einasto--like halo is adopted. 
This low-mass extrapolation depends on whether the outer RC decline remains as steep as inferred here. \textit{Gaia} DR4 will provide an important test of this behavior. The present analysis shows that the declining outer RC and the associated Einasto model are not removed when the tracer density is measured from the same stellar sample and the main outer disk systematic terms are propagated. It also shows that the inferred total mass depends strongly on the assumed halo profile, and that NFW or gNFW extrapolations can overestimate the MW mass when the outer RC declines significantly. A meaningful comparison with large radius tracers therefore requires explicit treatment of their dynamical state.

The mass estimates above assume Newtonian gravity and a standard baryon-DM decomposition. Alternative interpretations based on modified Newtonian dynamics (MOND) or the empirical radial acceleration relation would affect the subsequent mass interpretation rather than the Jeans reconstruction of the RC \citep{Milgrom1983a, McGaughRAR2016, Klacka2026}. A systematic comparison of such frameworks is beyond the scope of this work, but the tabulated RC and the adopted baryonic model parameters allow independent tests.


\subsection{Missing or excess baryon problem?}
\label{sec:missing_baryon}

The missing baryon problem refers to the difference between the baryon fractions measured in bound structures and the universal value \citep[e.g.][]{McGaugh2010}. The cosmic baryon fraction is well constrained by cosmic microwave background (CMB) cosmology,
$f_{b,\rm cos}\equiv\Omega_b/\Omega_m\simeq0.156$ \citep{Planck2020}.
Recent work has shown that this problem depends strongly on gas phase and radius. Fast Radio Burst (FRB) dispersion measures indicate that a large fraction of baryons resides in diffuse ionized gas in the IGM \citep{Macquart2020,Connor2025}. On galaxy halo scales, eROSITA stacking analyses suggest that the hot CGM traced within $R_{\rm vir}$ remains below the cosmic baryon fraction, while baryon closure may be approached only at several virial radii \citep{Zhang2024}. Thermal Sunyaev Zeldovich measurements also show that the inferred CGM baryon content depends on the assumed gas thermal state \citep{Das2023}.

In the low-mass Einasto extrapolation favored by the present RC fit, the baryon accounting differs from this interpretation. For the fiducial Einasto model, we find $M_\mathrm{tot}=2.01\times10^{11}\,M_\odot$, with upper limits of $2.17\times10^{11}\,M_\odot$ at $1\sigma$ and $2.53\times10^{11}\,M_\odot$ at $3\sigma$. 
The baryonic model adopted here already contains about $6\times10^{10}\,M_\odot$. This gives $f_b\simeq0.24$ to $0.31$, before adding any extended CGM component. Thus, in the low mass Einasto extrapolation favored by the present RC, the MW does not appear to have a missing baryon problem. It instead shows a possible baryon excess tension.

This tension is relevant because recent MW studies support the presence of warm and hot ionized gas, but do not require an extremely massive CGM. The eROSITA O~VIII analysis suggests that much of the soft X--ray emission at intermediate and high latitude comes from a flattened disk related component, although a more extended spherical halo may still contribute to absorption and to the CGM baryon content \citep{Locatelli2024}. FRB based constraints also limit the MW halo electron column and CGM mass, with halo DM contributions of tens to at most $\sim10^2\,{\rm pc\,cm^{-3}}$ and a conservative upper limit of $<10^{11}\,M_\odot$ from FRB 20220319D \citep{ProchaskaZheng2019,Cook2023,Ravi2025}. Any substantial additional baryonic reservoir would therefore further increase the baryon fraction tension in our low mass model.

For a canonical $M_{200}\sim10^{12}\,M_\odot$ halo, the baryon problem is primarily one of locating the missing baryons. For the low-mass halo considered here, the relevant consistency check is whether the halo mass, known baryonic mass, and CGM constraints can be reconciled. A similar tension may also arise in M31. \citet{Hammer2025} derived a dark matter mass of $M_{200}=2.95\times10^{11}\,M_\odot$ and a total mass of $4.5\times10^{11}\,M_\odot$ within 137 kpc from hydrodynamical simulations matched to its recent merger history, implying a baryon fraction of 0.32. Although M31 is dynamically more complex than the MW, this comparison suggests that the baryon content of massive spirals may need to be reconsidered when low halo masses and equilibrium assumptions are jointly examined.

\section{Conclusions}
\label{sec:conclusions}

In this work, we derived a new MW RC from \textit{Gaia} DR3 using low $\alpha$ upper RGB stars and the radial Jeans equation. The tracer sample was selected to reduce contamination from counter rotating stars, strongly non circular orbits, and other kinematic outliers. This makes the sample better suited to the assumptions of a Jeans analysis. A key improvement is that the tracer density profile was measured from the same stars used for the kinematic moments, rather than being arbitrarily fixed from an external scale length. We also included the cross term directly in the fiducial RC and estimated the main systematic uncertainties from Solar parameters, radial gradient terms, the cross term, and spatial splitting terms.

The tracer density profile is better described by a double exponential than by a single exponential profile. This changes the density term in the Jeans equation, especially in the inner disk. The resulting RC is nearly flat in the inner disk and then declines with radius. The decrease becomes clear beyond $R\sim15$ to $16$ kpc, and the fitted outer slope is close to Keplerian over the adopted interval beyond $R\sim16$ kpc. The systematic uncertainties increase in the outer disk, mainly because of disk azimuthal and vertical asymmetries that account for the observed perturbations. However, these uncertainties do not remove the declining trend. The cross term remains a small contribution compared with the large scale asymmetries traced by the splitting tests.

Our analysis also addresses the concern that the outer decline could be caused by an incorrect tracer density profile. Over the radial range where the density can be measured most reliably, $12.5<R<21$ kpc, we do not find evidence for a sharp truncation of the RGB tracer population. The density profile is instead smooth and extended. A sharp truncation within this range is therefore unlikely to be the dominant cause of the observed outer RC decline, although a further break cannot be excluded.

The mass modeling favors a low MW mass for which the DM halo is well described by an Einasto profile. The best fit gives $M_\mathrm{dyn}= 2.01^{+0.10}_{-0.08} \times10^{11} M_\odot$. This low mass follows from the slow growth of the enclosed mass over the radial range directly constrained by the RC. When an NFW halo is imposed, the best fitting mass increases to $6.65\times10^{11} M_\odot$, but only with a $\chi^2$ probability of $p=0.01$. This shows that the inferred virial mass depends strongly on the adopted halo profile.

The extrapolation from the measured disk RC to $M_{200}$ should therefore be interpreted with care. The RC directly constrains the Galactic gravitational field only out to the last measured point, whereas $M_{200}$ depends on the halo structure at much larger radii. In this sense, our result supports a low mass extrapolation for an Einasto--like halo, but it does not provide a model-independent measurement of the virial mass. Reconciling the declining outer RC with a much more massive halo would require substantial mass growth beyond the last disk point without producing a strong signature in the measured RC.

After measuring the tracer density from the RC sample itself, including the cross term, and propagating the main outer disk asymmetry terms, the MW disk RC remains declining at large radii. The dominant remaining uncertainties are linked to the imperfect equilibrium and non-axisymmetry of the outer disk. A deeper understanding of the dynamical evolution of the MW disk and halo is thus required, not only for the present analysis but also for mass estimates from tracers at larger radii, such as dwarf galaxies, whose equilibrium must be established for using their orbital motions to reliably probe the Galactic potential.

\begin{acknowledgments}
This work was supported by the National Natural Science Foundation of China (NSFC; Grant No. 12588202). Y.J. acknowledges support from the Shuimu Tsinghua Scholar Program of Tsinghua University. This work has made use of data from the European Space Agency mission \textit{Gaia}, processed by the \textit{Gaia} Data Processing and Analysis Consortium. This work also uses data from the Apache Point Observatory Galactic Evolution Experiment, which is part of the Sloan Digital Sky Survey IV.
\end{acknowledgments}

\software{astropy \citep{Astropy2013,Astropy2018,Astropy2022},  SciPy \citep{Scipy2020}, galpy \citep{Galpy2015}, emcee \citep{emcee2013}
          }


\appendix

\section{Effect of the vertical selection}
\label{app:systematic_z}

\begin{figure}[htb!]
\centering
\includegraphics[width=\columnwidth]{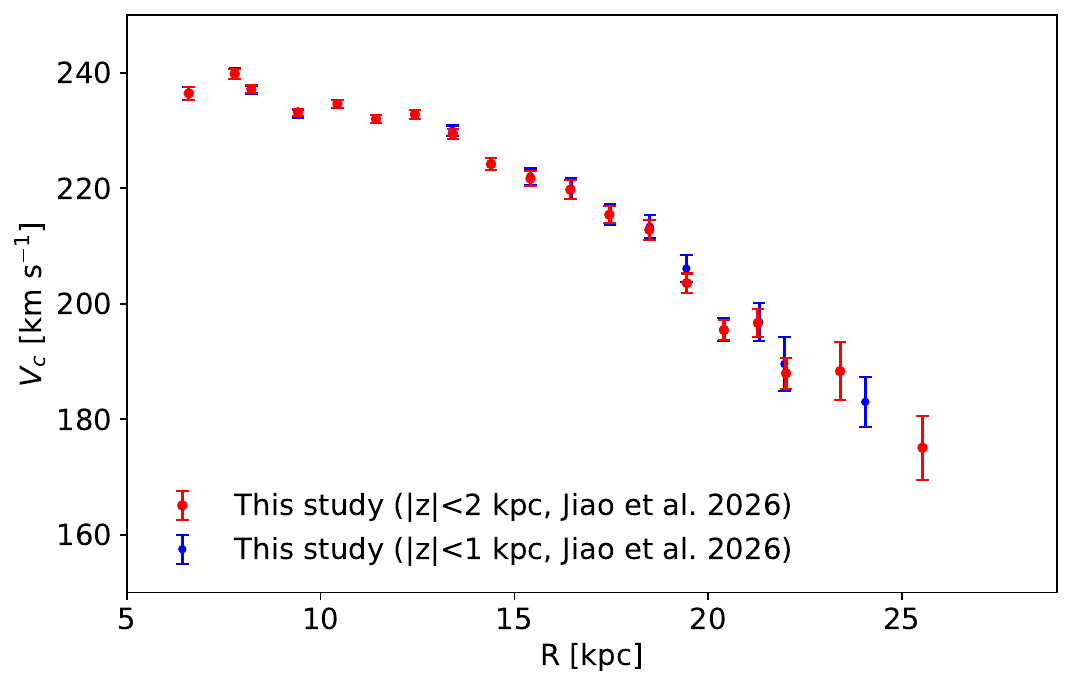}
\caption{Comparison between the fiducial RC measured with $|z|<2\,\mathrm{kpc}$ and the stricter vertical selection $|z|<1\,\mathrm{kpc}$. The two measurements agree over their common radial range. The error bars show only statistical uncertainties. The $|z|<1\,\mathrm{kpc}$ selection is used only as a consistency check, because it reduces the number of outer disk tracers and can be more sensitive to the detailed warped midplane geometry.}
\label{fig:rc_z_comparison}
\end{figure}

In Section~\ref{sec:data}, we adopt a fiducial vertical selection of $|z|<2\,\mathrm{kpc}$. This choice keeps the outer disk coverage of the original wedge sample while avoiding an overly restrictive cut around the midplane of a warped and flaring disk. To test whether this choice affects the inferred RC, we repeat the measurement using a stricter selection of $|z|<1\,\mathrm{kpc}$.

Figure~\ref{fig:rc_z_comparison} compares the resulting RC with the fiducial one. The two curves agree over their common radial range. The uncertainties shown in this comparison are statistical only and do not include the systematic uncertainty estimate from Section~\ref{sec:systematics}. The stricter cut also reduces the number of available tracers in the outer disk, and is not adopted as the fiducial selection.

\section{KS $p$ values for the luminosity function comparison}
\label{app:ks_p_values}

\begin{figure}[htb!]
  \centering
  \includegraphics[width=\columnwidth]{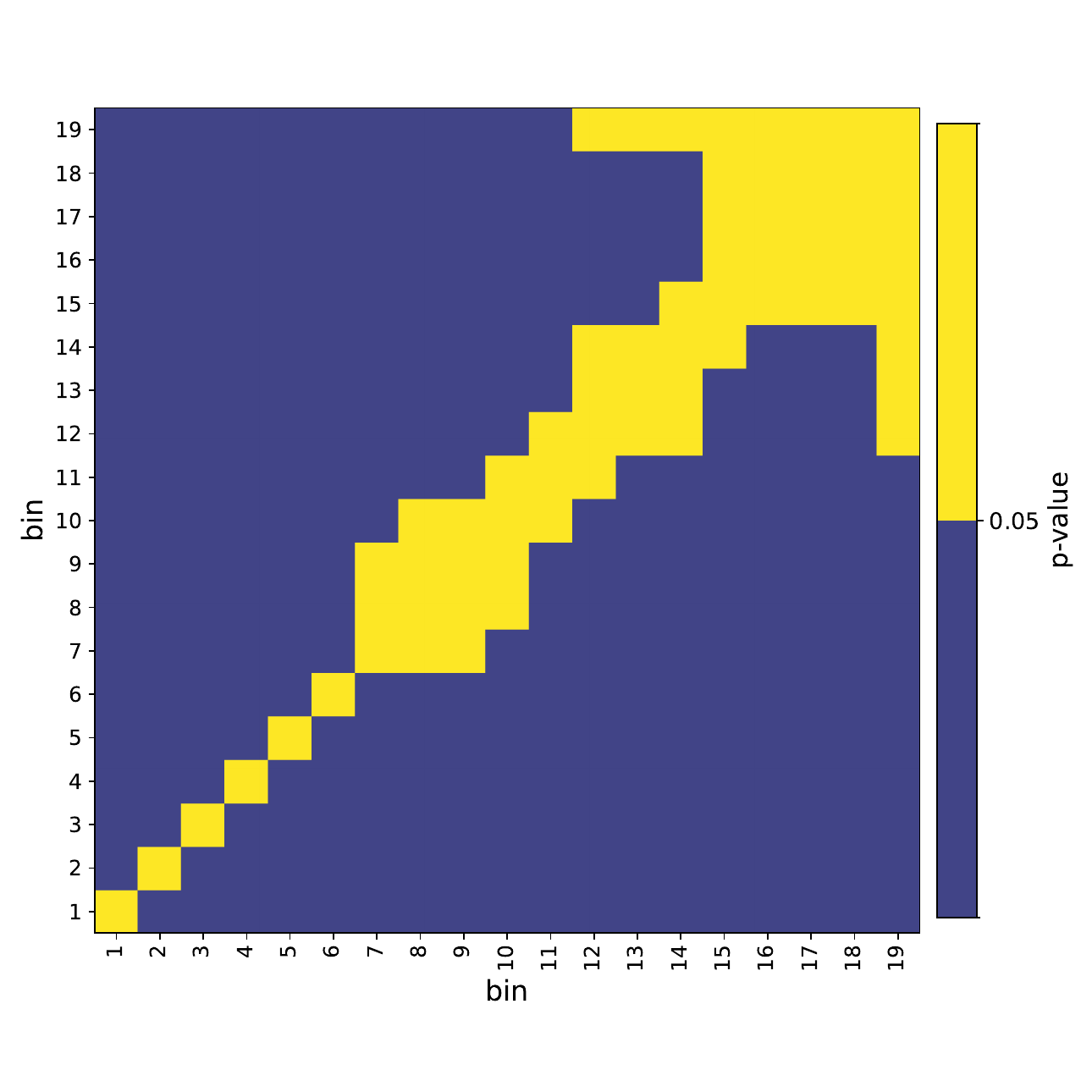}
  \caption{Pairwise KS $p$ value categories for the $H_{\rm abs}$ distributions in different radial bins. The categories are defined relative to $p=0.05$. Because the $p$ value depends on sample size, this figure is used only as a qualitative check of the fitted radial range selected from the KS $D$ statistic.}
  \label{fig:ks_p_values}
\end{figure}

Figure~\ref{fig:ks_p_values} shows the pairwise KS $p$ value categories for the same $H_{\rm abs}$ distributions discussed in Section~\ref{sec:density_range}. The $p$ value depends on both the distribution difference and the sample sizes of the two bins. We therefore do not use it as the primary criterion for selecting the fitted radial range. Instead, it is used only as a qualitative check on the KS $D$ statistic shown in Fig.~\ref{fig:ks_pair}. The adopted interval, $12.5<R<21\,\mathrm{kpc}$, includes the main contiguous region where neighbouring or nearby bins often have non rejected pairwise comparisons.

\section{Local gradient diagnostics for the systematic uncertainty estimates}
\label{app:gradient_diagnostics}

This appendix presents the local MCMC estimates used to assess the systematic uncertainties associated with the gradient terms in the Jeans equation. These estimates are diagnostic tests. They are not used to replace the fiducial smooth profiles or the adopted cross term prescription in the main RC calculation.

\subsection{Cross term}
\label{app:cross_term_bin}

\begin{figure*}[htb!]
\centering
\includegraphics[width=\textwidth]{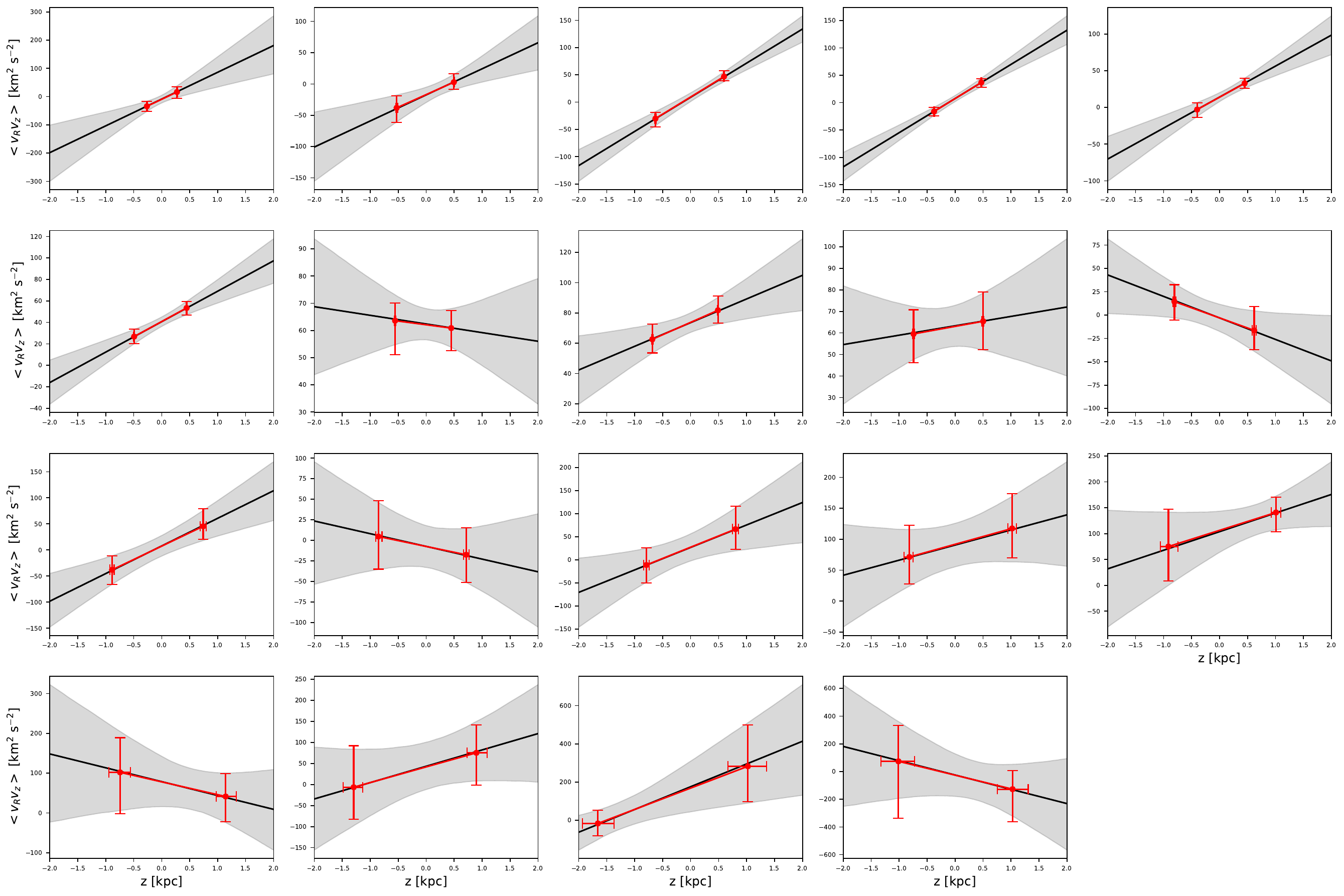}
\caption{Estimation of the vertical gradient of $\langle v_R v_z \rangle$ in individual radial bins. Red points show the measurements for the subsamples above and below the Galactic plane. The black lines and gray shaded regions show the median and $68\%$ credible interval of the MCMC linear fits.}
\label{fig:cross_term_bin}
\end{figure*}

Figure~\ref{fig:cross_term_bin} shows the bin by bin estimates of the vertical gradient of $\langle v_R v_z\rangle$. In each radial bin, the sample is divided into stars above and below the Galactic midplane. The two measurements of $\langle v_R v_z\rangle$ are then fitted as a linear function of $z$ with an MCMC procedure. These fits provide the local estimates shown in Fig.~\ref{fig:cross_term_all}.

The bin by bin fits show a rapid change in the cross term in the inner Galaxy. From $R\simeq6$ to $10\,\mathrm{kpc}$, the inferred gradients decrease with radius, possibly reflecting inner non-axisymmetric perturbations. At larger radii, the measurements are flatter on average, but the individual estimates become increasingly uncertain as the number of tracers decreases. The absence of a clear radial trend beyond $R\simeq10\,\mathrm{kpc}$, together with the large outer uncertainties, motivates the use of an inverse variance weighted average for the fiducial RC calculation. The local measurements are therefore used mainly to quantify the systematic uncertainty associated with the cross term, rather than as direct noisy corrections in each outer radial bin.

\subsection{Radial gradient terms}
\label{app:gradient_systematics}

We also use local MCMC estimates to test the sensitivity of the two radial gradient terms,
$\partial\ln\nu/\partial\ln R$ and
$\partial\ln\langle v_R^2\rangle/\partial\ln R$.
We construct an alternative set of radial bins with a width of $1\,\mathrm{kpc}$. The bin centres are placed at the weighted mean radii of the fiducial bins, corresponding approximately to a half bin shift. Local logarithmic slopes are then estimated from adjacent rebinned measurements with an MCMC linear fit in logarithmic space.

\subsubsection{Tracer density gradient}
\label{app:density_gradient}

\begin{figure}[htb!]
\centering
\includegraphics[width=0.9\columnwidth]{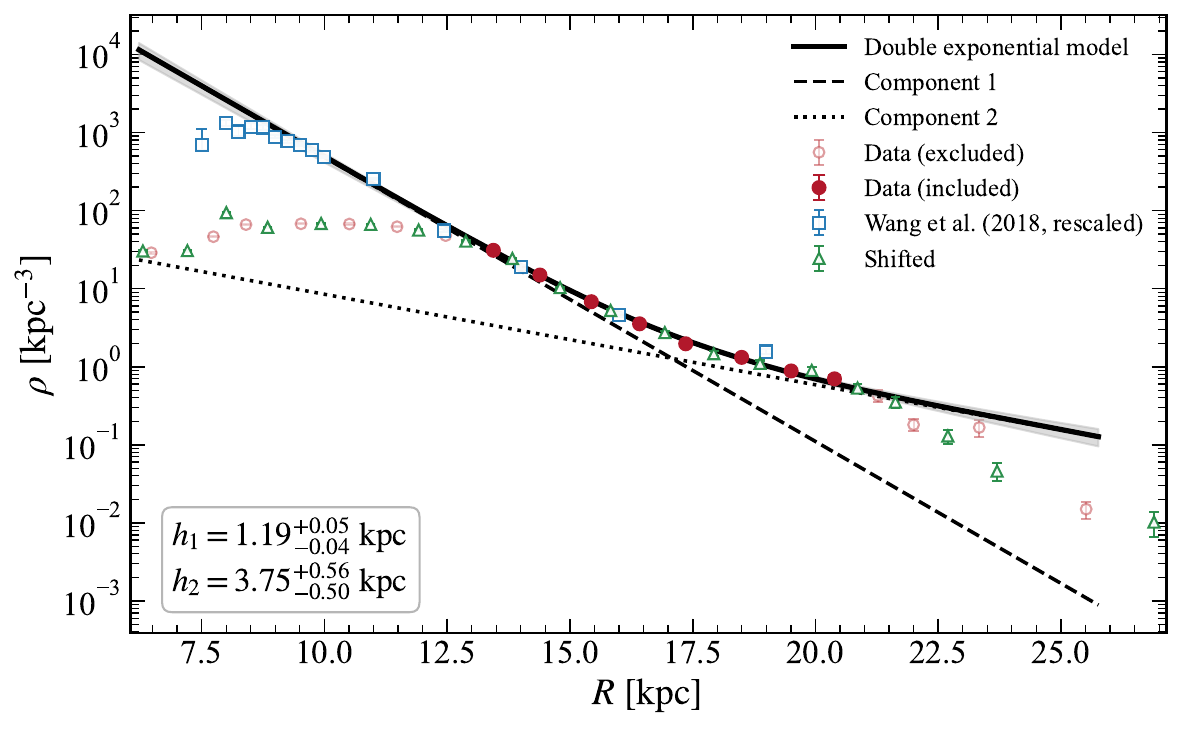}
\caption{Tracer density profile used to test the sensitivity of the density gradient to the adopted radial binning. Same as Fig.~\ref{fig:fig_dens}. Green triangles show the shifted density measurements used only for the local gradient test.}
\label{fig:density_shifted_profile}
\end{figure}

Figure~\ref{fig:density_shifted_profile} compares the shifted density measurements with the fiducial double exponential profile. The shifted points follow the same global trend over the validated density fitting range. At smaller and larger radii, they deviate more strongly, as expected from magnitude selection effects and small number statistics.

\begin{figure}[htb!]
\centering
\includegraphics[width=0.9\columnwidth]{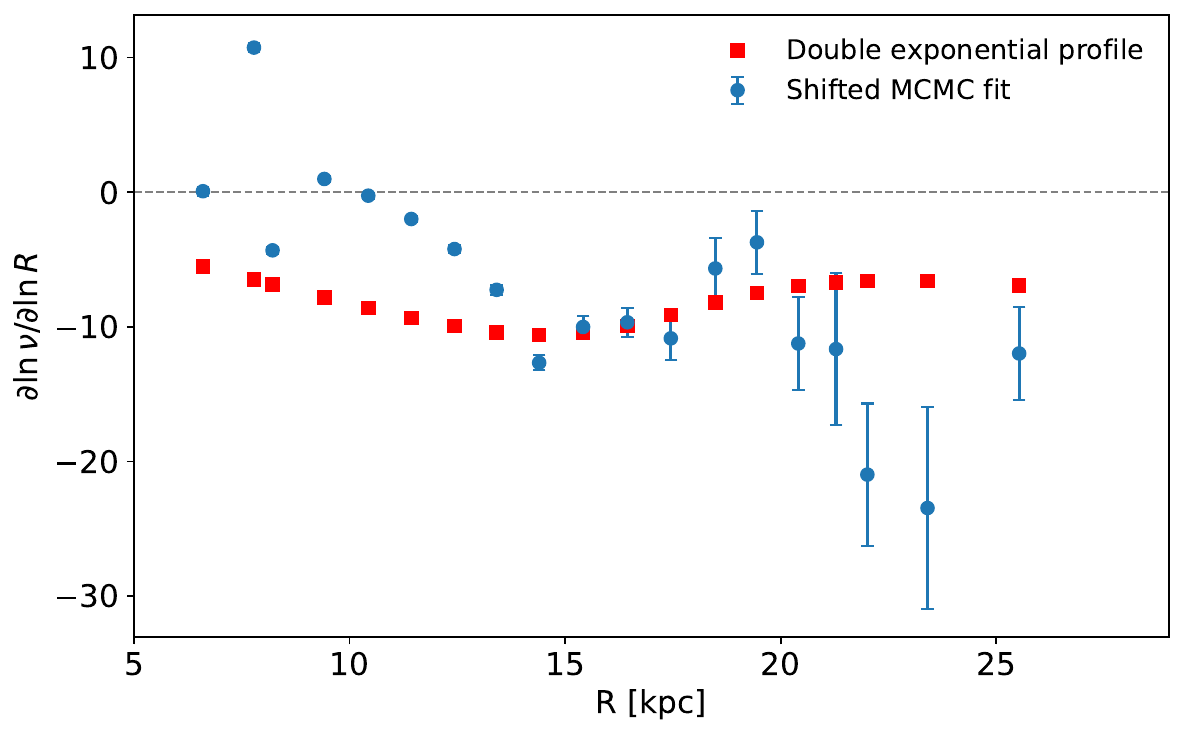}
\caption{Local estimates of the tracer density gradient. Red squares show the gradient of the fiducial double exponential profile. Blue points show the local MCMC estimates obtained from adjacent shifted density measurements. The gray dashed line marks zero. These estimates are used only to evaluate the systematic uncertainty associated with the density gradient.}
\label{fig:density_gradient_local}
\end{figure}

\begin{figure*}[htb!]
\centering
\includegraphics[width=\textwidth]{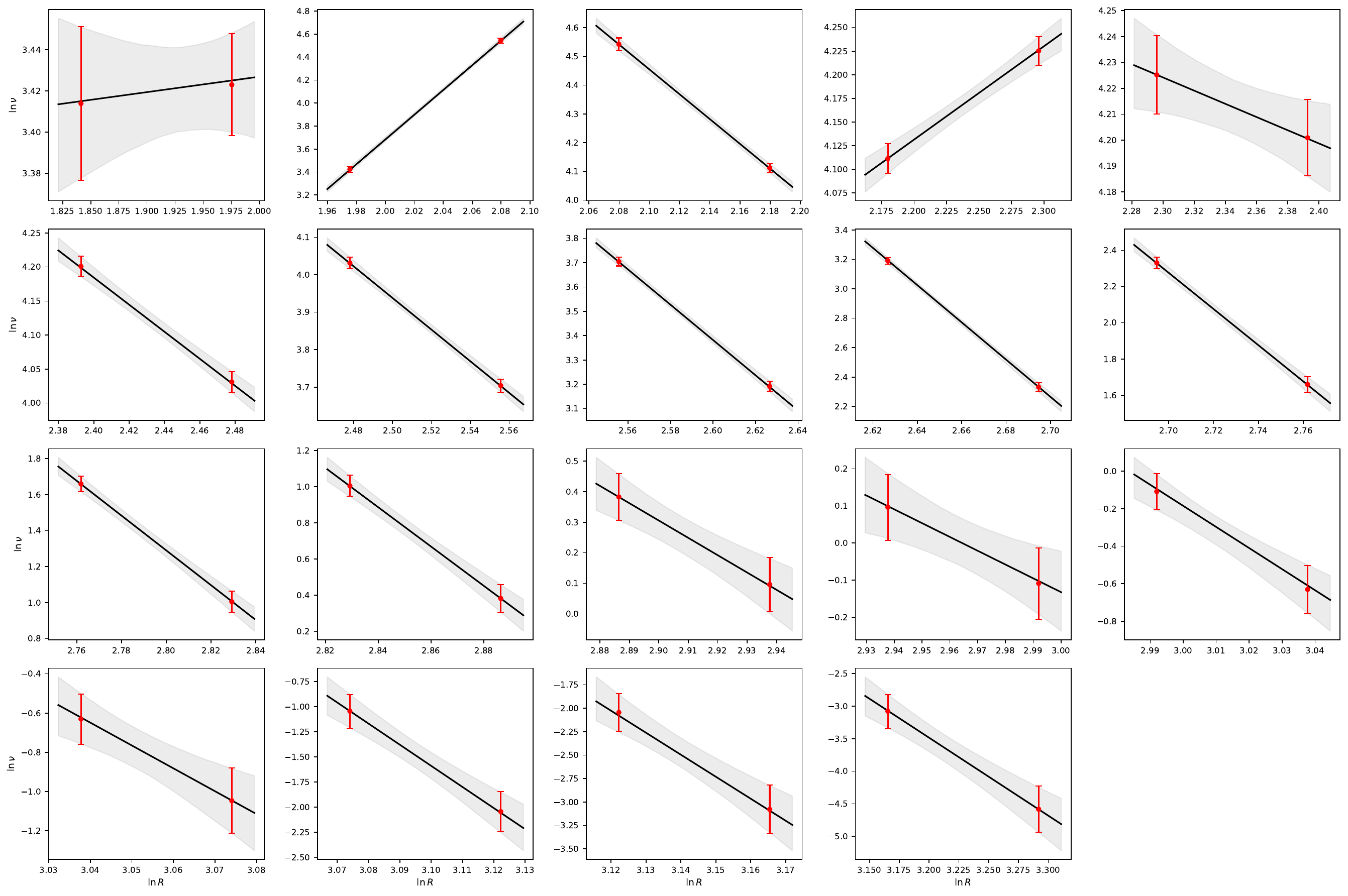}
\caption{MCMC fits used to estimate the local tracer density gradients. Each panel shows the linear fit in the $\ln\nu$ versus $\ln R$ plane for one local radial interval. Red points show the two adjacent shifted measurements used in the fit. The black line and gray shaded region show the median relation and the $68\%$ credible interval.}
\label{fig:density_gradient_mcmc}
\end{figure*}

For each local interval, we fit a relation between $\ln\nu$ and $\ln R$ using the two adjacent shifted density measurements. The slope gives the local estimate of $\partial\ln\nu/\partial\ln R$. Figure~\ref{fig:density_gradient_local} compares these estimates with the gradient of the fiducial double exponential profile, and Fig.~\ref{fig:density_gradient_mcmc} shows the corresponding MCMC fits.

The local density gradients agree reasonably well with the fiducial model within the validated density fitting range. Outside this range, the local slopes are less reliable because the density measurements are more sensitive to selection effects and small number statistics. We therefore use the local density gradients directly only within the fitted interval. Outside this range, the density gradient systematic is extrapolated from representative values measured inside the fitted region, as described in Section~\ref{sec:systematics}.

\subsubsection{Radial gradient of $\langle v_R^2\rangle$}
\label{app:vr2_gradient_local}

\begin{figure}[htb!]
\centering
\includegraphics[width=0.9\columnwidth]{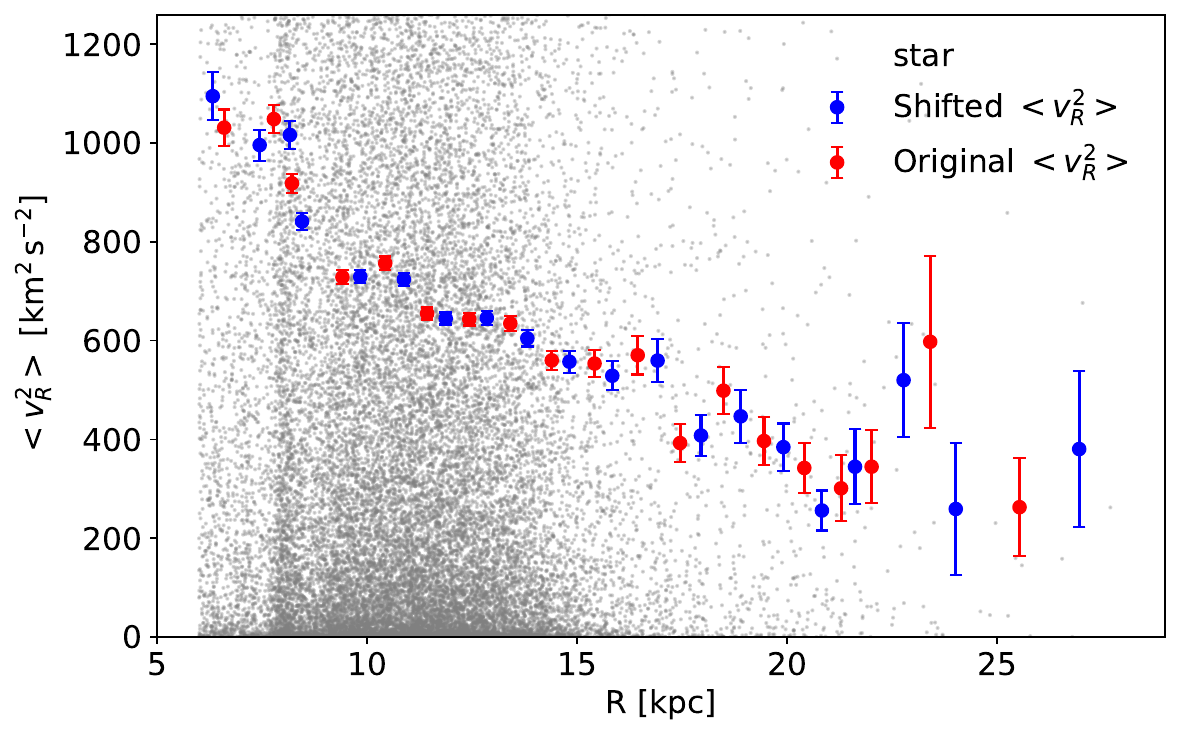}
\caption{Radial profile of $\langle v_R^2\rangle$ for the fiducial and shifted radial bins. Gray points show individual stars. Red points show the fiducial binned measurements, while blue points show the shifted measurements used for the local gradient test.}
\label{fig:vr2_shifted_profile}
\end{figure}

We apply the same shifted binning test to the radial second moment. Figure~\ref{fig:vr2_shifted_profile} compares the original and shifted measurements of $\langle v_R^2\rangle$. The shifted measurements follow the same radial decline as the fiducial profile over most of the disk. At the largest radii, the uncertainties increase because the number of stars per bin becomes small.

\begin{figure}[htb!]
\centering
\includegraphics[width=0.9\columnwidth]{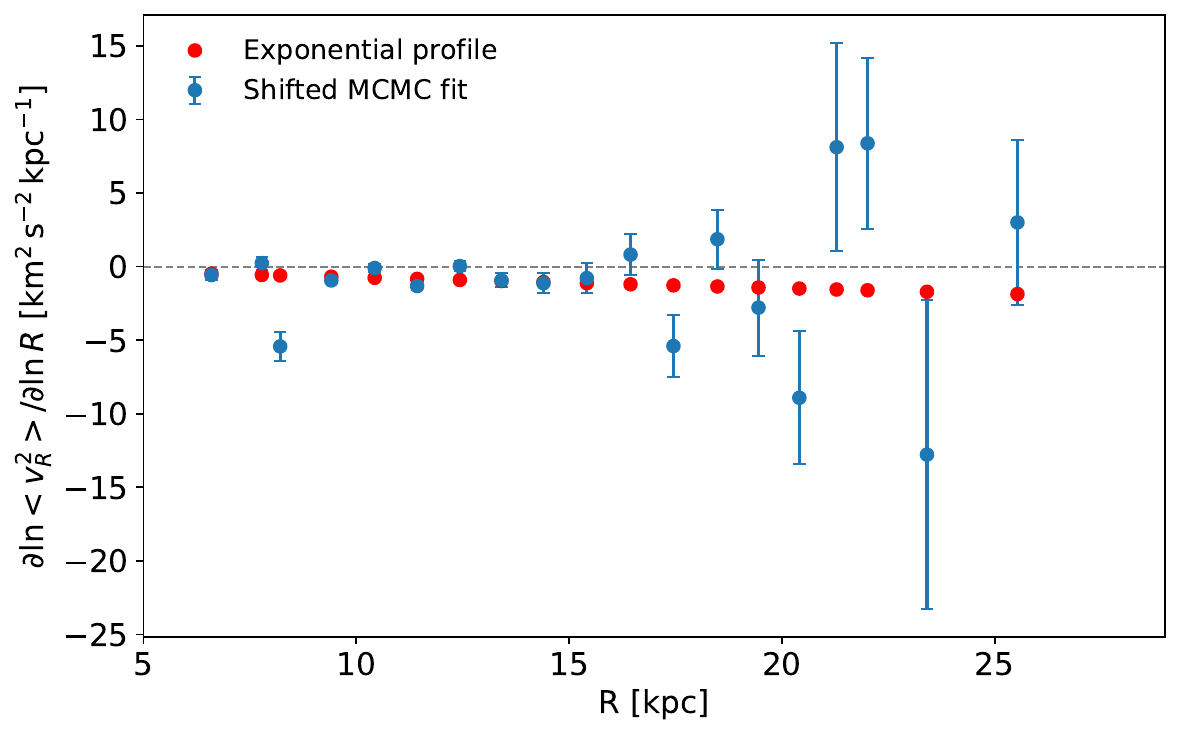}
\caption{Local estimates of the radial gradient of $\langle v_R^2\rangle$. Red points show the gradient of the fiducial exponential profile. Blue points show the local MCMC estimates obtained from adjacent shifted measurements. The gray dashed line marks zero. These estimates are used to quantify the systematic uncertainty associated with the radial second moment gradient.}
\label{fig:vr2_gradient_local}
\end{figure}

\begin{figure*}[htb!]
\centering
\includegraphics[width=\textwidth]{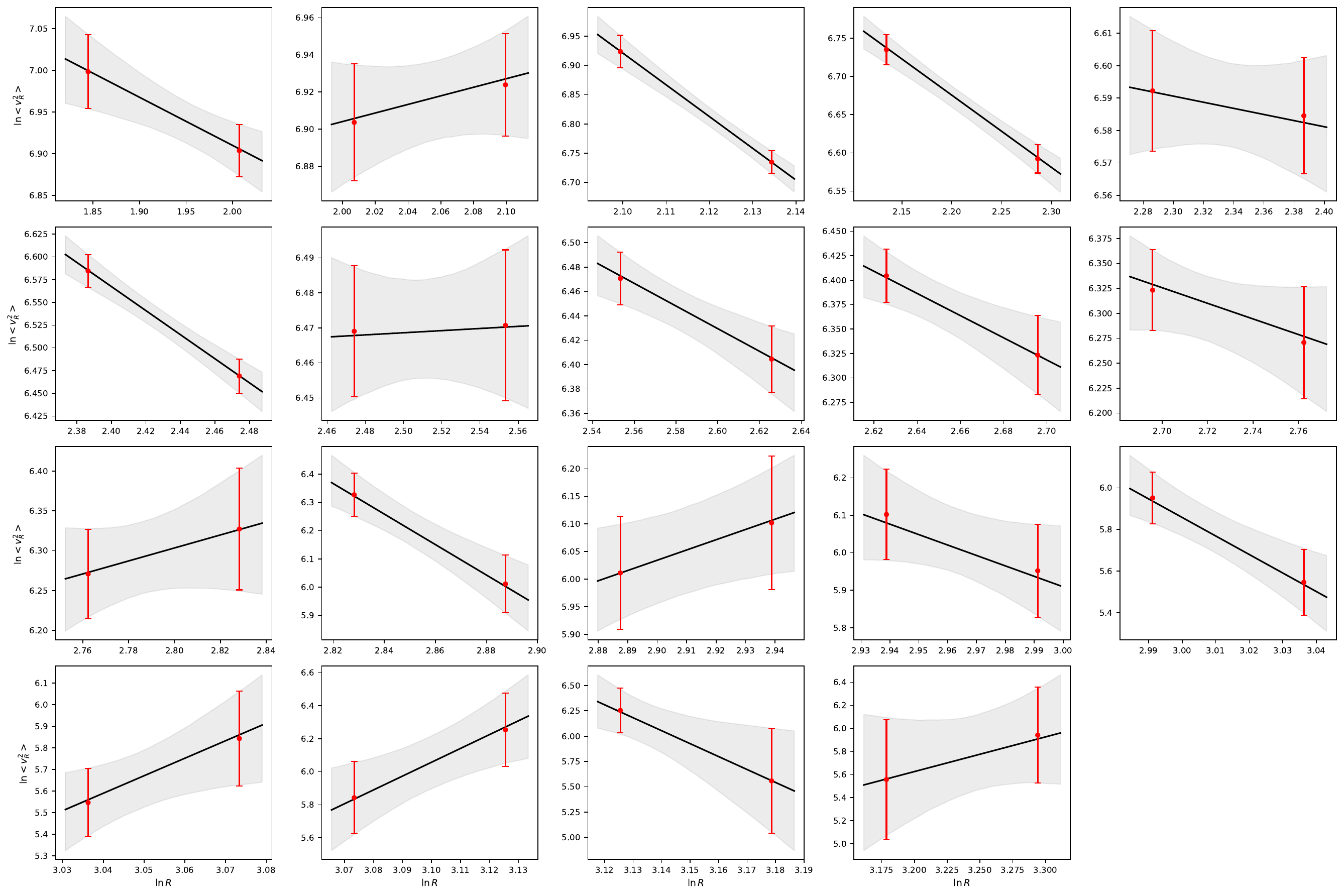}
\caption{MCMC fits used to estimate the local gradients of $\langle v_R^2\rangle$. Each panel shows the linear fit in the $\ln\langle v_R^2\rangle$ versus $\ln R$ plane for one local radial interval. Red points show the two adjacent shifted measurements used in the fit. The black line and gray shaded region show the median relation and the $68\%$ credible interval.}
\label{fig:vr2_gradient_mcmc}
\end{figure*}

For each local interval, we fit a relation between $\ln\langle v_R^2\rangle$ and $\ln R$. The slope gives the local estimate of $\partial\ln\langle v_R^2\rangle/\partial\ln R$. Figure~\ref{fig:vr2_gradient_local} compares these estimates with the gradient predicted by the fiducial exponential profile, and Fig.~\ref{fig:vr2_gradient_mcmc} shows the individual MCMC fits.

Unlike the density gradient, the local estimates of $\partial\ln\langle v_R^2\rangle/\partial\ln R$ can be obtained over the full radial range of the RC measurement. We propagate the difference between these local estimates and the fiducial exponential gradient through the Jeans equation. The resulting change in the RC is used as the systematic uncertainty associated with this term.

\section{Diagnostics of vertical and azimuthal structure in the selected sample}
\label{app:vphi_r_diagnostics}

\begin{figure}[htb!]
\centering
\includegraphics[width=0.95\columnwidth]{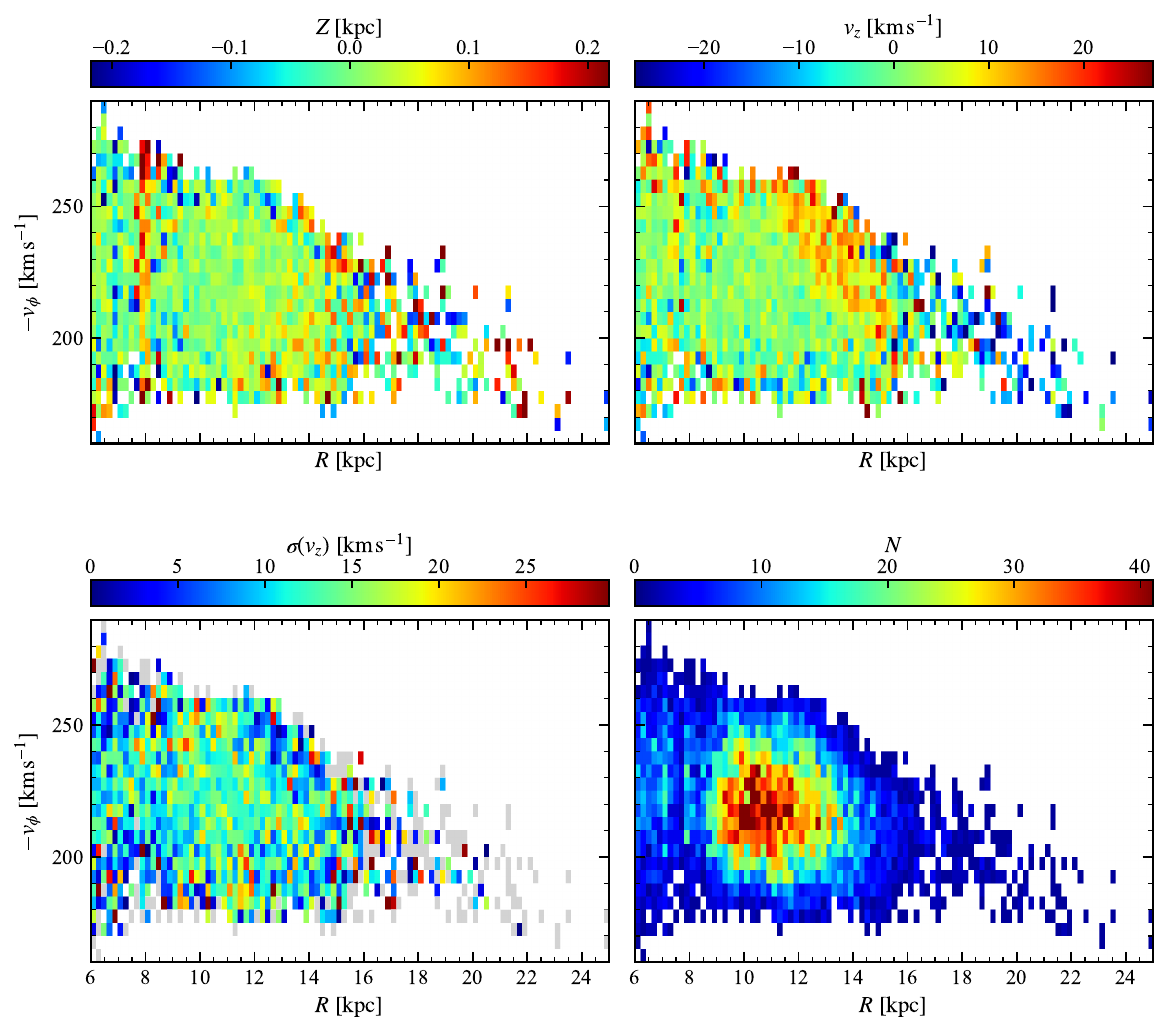}
\caption{Diagnostic maps in the $R$ and $-v_\phi$ plane for the near plane subsample with $|Z|<0.25\,\mathrm{kpc}$. The four panels show the same bins coloured by the median height $Z$ (upper left), the median vertical velocity $v_z$ (upper right), the robust vertical velocity scatter $1.5\,{\rm MAD}(v_z)$ (lower left), and the number of stars $N$ (lower right). The lower left panel uses gray pixels for bins containing only one star, for which the MAD cannot be estimated.}
\label{fig:vphi_r_nearplane}
\end{figure}

We compare the kinematic structure of our selected tracers with the outer disk features reported by \citet{Antoja2021} and \citet{Drimmel2023}. For this purpose, we examine the distribution of stars in the $R$ and $-v_\phi$ plane following the same vertical cuts as done in Fig. 14 of \citet{Antoja2021}. In each bin, we compute the median height $Z$, the median vertical velocity $v_z$, the robust vertical velocity scatter $1.5\,{\rm MAD}(v_z)$, and the number of stars. This diagnostic is not used in the RC calculation. It is intended to check whether the selected sample contains a strong vertical kinematic separation, such as a branch in which stars below the plane rotate faster than those above it.

Figure~\ref{fig:vphi_r_nearplane} shows the near plane subsample with $|Z|<0.25\,\mathrm{kpc}$. This selection is useful for comparison with the anticentre maps of \citet{Antoja2021}. We do not find a clear branch in which stars below the plane rotate systematically faster than stars above the plane. The $v_z$ panel shows mild structure, indicating that the sample is not perfectly phase mixed, but the signal is weaker than the outer disk features seen in broader \textit{Gaia} samples.

\begin{figure}[htb!]
\centering
\includegraphics[width=0.95\columnwidth]{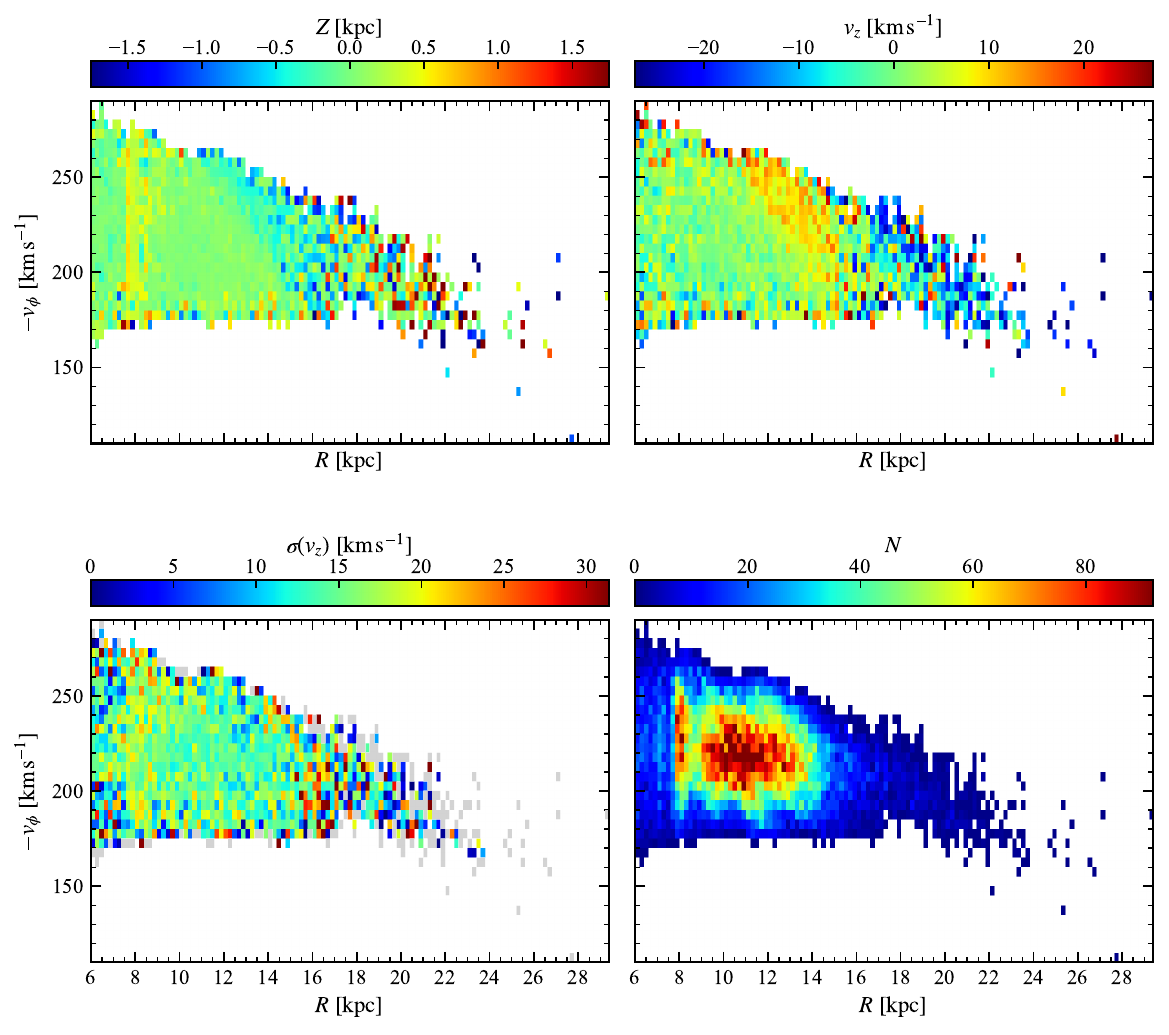}
\caption{Same as Fig.~\ref{fig:vphi_r_nearplane}, but for the full selected star sample used in the RC analysis.
}
\label{fig:vphi_r_fullsample}
\end{figure}

Figure~\ref{fig:vphi_r_fullsample} shows the same diagnostic for the full sample used in the RC analysis. The larger vertical range increases the number of stars and reveals the expected broadening of the vertical structure. The same conclusion holds: the selected tracer sample does not show a strong below plane, faster rotating branch. This supports the importance of the kinematic selection used in this work. Residual vertical and azimuthal asymmetries are still propagated through the splitting tests described in Section~\ref{sec:systematics}.

\bibliography{ref}{}
\bibliographystyle{aasjournalv7}


\end{CJK*}
\end{document}